\documentclass[aps,prl,twocolumn,superscriptaddress]{revtex4}

\usepackage{amsmath,bm}
\usepackage{mathrsfs}
\usepackage{amsfonts}
\usepackage{graphicx}

\usepackage{placeins}
\usepackage{setspace} 
\usepackage{epstopdf}
\usepackage{dcolumn}
\usepackage{amsmath}
\usepackage{epsfig}
\usepackage{indentfirst}
\usepackage{psfrag}
\usepackage{subfigure}
\usepackage{float}
\usepackage{amssymb}
\usepackage{color}
\usepackage{units} 
\usepackage{graphicx}
\usepackage{dcolumn}
\usepackage{bm}
\usepackage{physics}
\usepackage{natbib}
\usepackage{float}

\usepackage{xr-hyper}     
\usepackage[backref=none,bookmarksnumbered=true,bookmarks=true,bookmarksopen=true,colorlinks=true,citecolor=blue,linkcolor=blue,filecolor=blue,anchorcolor=green,urlcolor=blue,unicode=false]{hyperref}
\usepackage{ulem}[normalem] 

\def\SNmethods{I}
\def\SNQPI{II}

\def\SMQPI-allYSR{IV}
\def\SNtheo{V}
\def\SNtheoo{VI}

\def\SFQPI0{S1}
\def\SFQPI{S2}
\def\SFQPI2{S3}
\def\SFOrbital{S4}
\def\SFQPI1{S5}

\newcommand{\rev}[1]{{\color{black} #1}}

\makeatletter
\newcommand\colorsout[1]{\bgroup \markoverwith{\textcolor{#1}{\rule[0.5ex]{2pt}{0.4pt}}}\ULon}

\makeatother

\newcommand{\bipd}{$\beta$-Bi$_2$Pd}
\def\ef{E$_F$}

\def\alfap{$\alpha_+$}

\def\ss1{S$_{\rm{1}}$}
\def\ss2{S$_{\rm{2}}$}
\def\b2{B$_{\rm{2}}$}
\def\b2{BS$_{\rm{2}}$}

\begin{document}

\title{Revealing inter-band  electron pairing in a superconductor with spin-orbit coupling}  


\author{Javier Zald\'ivar}   \thanks{These two authors contributed equally}
  \affiliation{CIC nanoGUNE-BRTA, 20018 Donostia-San Sebasti\'an, Spain}

\author{Jon Ortuzar}  \thanks{These two authors contributed equally}
  \affiliation{CIC nanoGUNE-BRTA, 20018 Donostia-San Sebasti\'an, Spain}

\author{Miguel Alvarado}
  \affiliation{ 
Instituto de Ciencia de Materiales de Madrid (ICMM), Sor Juana In{\'e}s de la Cruz 3, 28049 Madrid, Spain}
  \affiliation{Departamento de Física Teórica de la Materia Condensada and Condensed Matter Physics Center (IFIMAC), Universidad Autónoma de Madrid, Madrid, Spain }

\author{Stefano Trivini}
  \affiliation{CIC nanoGUNE-BRTA, 20018 Donostia-San Sebasti\'an, Spain}

\author{Julie Baumard}
    \affiliation{Centro de Física de Materiales (CFM-MPC), Centro Mixto CSIC-UPV/EHU, 20018 San Sebastián, Spain}
  
\author{Carmen Rubio-Verd\'u }
  \affiliation{ICFO, 08860 Castelldefels, Barcelona, Spain}

\author{Edwin Herrera}
  \affiliation{Departamento de Física Teórica de la Materia Condensada and Condensed Matter Physics Center (IFIMAC), Universidad Autónoma de Madrid, Madrid, Spain }
  
\author{Hermann Suderow}
  \affiliation{Departamento de Física Teórica de la Materia Condensada and Condensed Matter Physics Center (IFIMAC), Universidad Autónoma de Madrid, Madrid, Spain }

\author{Alfredo Levy Yeyati}
  \affiliation{Departamento de Física Teórica de la Materia Condensada and Condensed Matter Physics Center (IFIMAC), Universidad Autónoma de Madrid, Madrid, Spain }

\author{F. Sebastian Bergeret}
    \affiliation{Centro de Física de Materiales (CFM-MPC), Centro Mixto CSIC-UPV/EHU, 20018 San Sebastián, Spain}
	\affiliation{Donostia International Physics Center (DIPC), 20018 Donostia-San Sebasti\'an, Spain}
  
\author{Jose Ignacio Pascual} 
  \affiliation{CIC nanoGUNE-BRTA, 20018 Donostia-San Sebasti\'an, Spain}
\affiliation{Ikerbasque, Basque Foundation for Science, Bilbao, Spain}

\begin{abstract}
\rev{Most superconducting mechanisms pair electrons within the same band, forming spin singlets. However, the discovery of multi-band superconductivity has opened new scenarios for pairing, particularly in systems with strong spin-orbit coupling. 
Here, we reveal inter-band pairing in the superconductor \bipd\ by mapping the amplitude of sub-gap Yu-Shiba-Rusinov (YSR) states around Vanadium adatoms deposited on its surface. The surface of \bipd\ is characterized by spin-helical-like bands near the Fermi level. Scanning tunneling spectroscopy reveals anisotropic YSR amplitude oscillations around the impurity, driven by spin-conserving Bogoliubov quasiparticle interference (BQPI). Analysis of the BQPI patterns at the YSR energy exposes inter-band pairing in this material. Interestingly, only a small subset of all possible inter-band scattering processes observed in the normal state contribute to the BQPI patterns. Combining experimental data and theory, we demonstrate that the observed band selectivity results from the hybridization of the band coupled with the impurity with other bands. Our findings reveal unconventional pairing mechanisms in \bipd\ and highlight the crucial role of spin-orbit interactions in their formation.
}

 \end{abstract}

\date{\today}
\maketitle

The complexity of electron pairing in superconductors underpins many novel states of matter with exotic properties. Due to the antisymmetric nature of the global wavefunction for two interacting fermions, the orbital character of the electronic bands determines the spin properties of the Cooper pair condensate. \rev{ As a consequence, pairs in the s-wave (p-wave) channel form a singlet (triplet) state \cite{Black-Schaffer2013}, leading to even-in-frequency superconductivity. The presence of spin-orbit coupling (SOC) adds further complexity to the pairing mechanisms by lifting spin degeneracies at the superconductor's surface and inducing helical-like bands. As illustrated in Fig.~1, electron pairing within spin-non-degenerate helical bands results in a mixture of spin-singlet and spin-triplet components in the superconducting condensate \cite{Gorkov2001,Kim2015}, which is essential for the emergence of topological superconductivity \cite{Sato2017}.

\begin{figure}[t]
  \includegraphics[width=0.99\columnwidth]{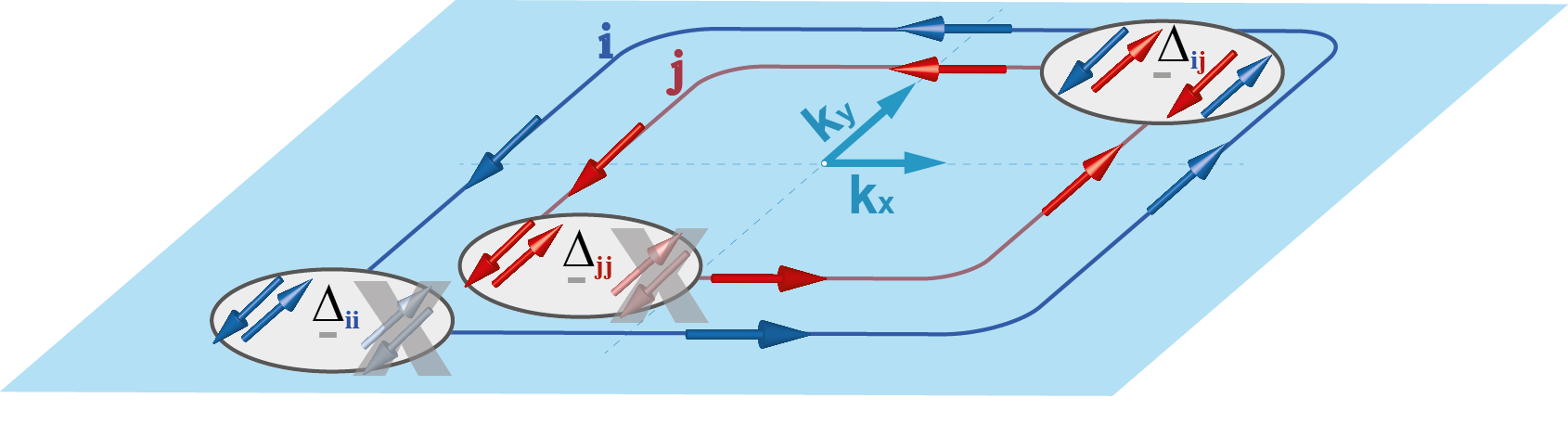}
  \caption{\rev{Illustration of superconducting pairing in a two-band superconductor with helical states. Intra-band singlet pairing is incomplete due to the lack of spin degenerate states for every band momentum. This results in a mix of singlet and triplet pairing components. Inter-band pairing leads to hybrid pairs formed by electrons from different bands, hence with a finite momentum.}  }  \label{intro0}
\end{figure}

The discovery of multi-band superconductors has broadened the landscape of possible mechanisms for pairing in the superconducting condensate. In such a scenario of helical bands as depicted in   Fig.~1, pairing of electrons from different bands can result in \textit{hybrid} pairs with  finite momentum \cite{Asaba2024}  or in the lift of the even symmetry of the intraband pairing, enabling odd symmetry pairing schemes  \cite{Black-Schaffer2013}. 
However, despite theoretical predictions, the observation of band-mixed pairing schemes remains challenging because their different orbital character also hinders band hybridization \cite{Sprau2017,Sharma2020}. }

\begin{figure*}[t]
  \includegraphics[width=0.9\textwidth]{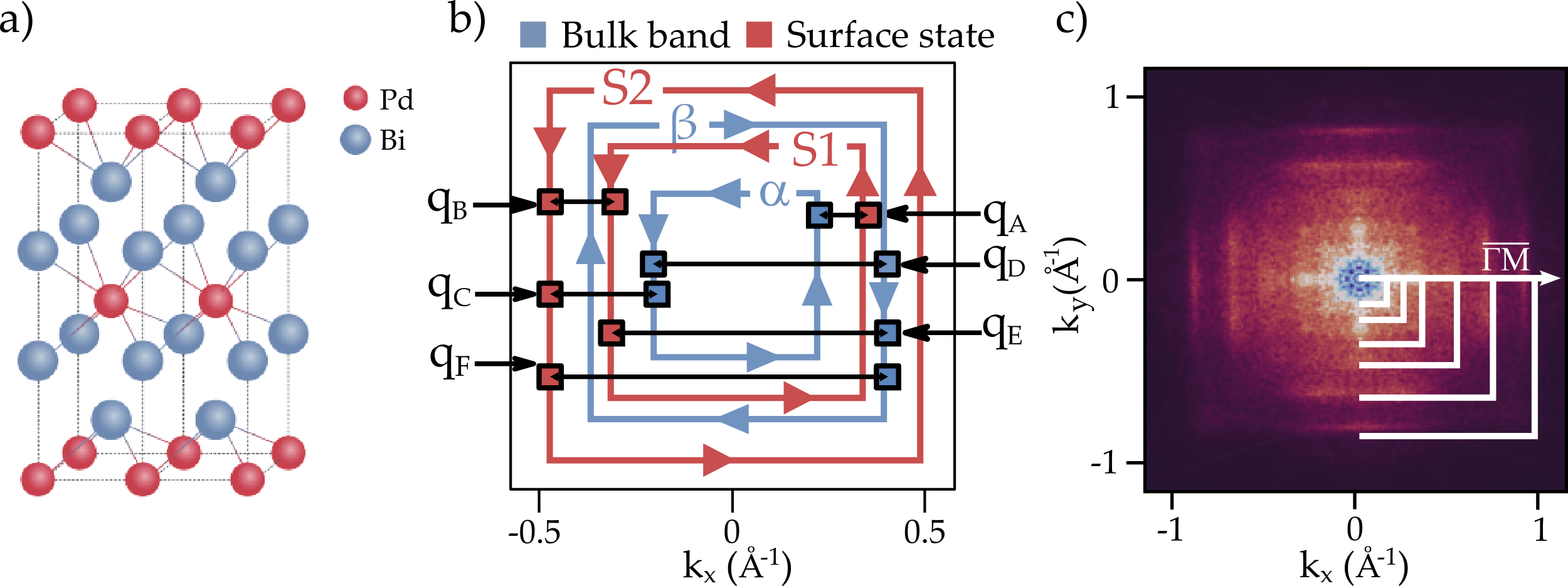}
  \caption{\textbf{The multiband superconductor \bipd:}  a) Structure of the layered \bipd\ superconductor. b) Schematic band structure on the \bipd\ surface, containing two spin non-degenerate surface bands, $S1$ and $S2$, and two projected bulk bands, $\alpha$ and $\beta$. The blue and red arrows in the contours indicate spin-polarization (simplified from refs.~\cite{Sakano2015,Iwaya2017}). The vectors, $q_i$, represent all possible spin-conserving scattering processes.  c) Fourier transformed dI/dV map measured with STM at 50 mV (details in Supplementary Note~\SNQPI) illustrating six contours (dashed lines) representing the six possible QPIs (blue dashed arrows) scattering vectors between the four bands. 
  }  \label{intro}
\end{figure*}

\rev{
In this letter, we present evidence of inter-band pairing in the conventional multi-band superconductor \bipd\ by analyzing sub-gap Bogoliubov quasiparticle interference (BQPI) patterns arising from magnetic impurities (Vanadium adatoms) on its Bi-terminated surface. Using a low-temperature scanning tunneling microscope (STM), we spatially map the amplitude of Yu-Shiba-Rusinov (YSR) states induced by vanadium adatoms. These states exhibit wave patterns along the high-symmetry directions of the substrate, with two characteristic wavelengths that provide insights into the sub-gap scattering of Bogoliubov quasiparticles.
The band structure of \bipd\ near the Fermi level consists of helical bands with a square-like Fermi contour \cite{Sakano2015}, spin-split due to strong SOC. As with quasiparticle interference (QPI) in the normal state \cite{petersen2000,pascual2004,pascual2011,Steinbrecher2013,Herrera2023}, spin conservation plays a crucial role in the scattering of Bogoliubov quasiparticles. Consequently, intra-band BQPI patterns are absent for helical bands, enabling the selective identification of inter-band scattering processes contributing to the BQPI patterns. Our results confirm the helical nature of the bands and reveal the coexistence of spin-singlet and spin-triplet superconducting pairing in \bipd.}

\textit{The superconducting alloy}:  \bipd\ is a layered material (Fig.~\ref{intro}a) with a transition temperature T$_c$=5.4~K and strong SOC.  Despite its complex band structure around the Fermi level, composed of several bulk bands and surface states with a square-like Fermi contours \cite{Imai2012,Sakano2015}, \bipd\  has a single superconducting gap of $\Delta_{sample}=\Delta_{tip}=$0.775 meV \cite{Herrera2015}. Owing to the strong SOC, electronic bands at the surface exhibit in-plane spin polarization. ARPES measurements and DFT simulations by Sakano et al. and Iwaya et al. \cite{Sakano2015,Iwaya2017} indicate that the \bipd\ surface host two helical surface bands, S1 and S2, illustrated in Fig.~\ref{intro}b, and several bulk bands projected on the surface, among which two of them, labeled B1 and B2, provide significant (spin-polarized) density of states (DoS) at the surface.

Using a low-temperature STM, we measured the normal-state QPI patterns of the as-cleaved Bi-terminated \bipd\ surface, mapping the differential conductance (dI/dV) at constant sample bias values well above the superconducting gap (see Supplementary Material, Sec. \SNmethods). The Fast Fourier Transformation (FFT) of these dI/dV maps reveals six squared contours [Fig. \ref{intro}(c)], indicating the presence of up to six scattering vectors, $\textbf{q}_i$, in agreement with refs.\cite{Sakano2015,Iwaya2017} [SM Sec. \SNQPI].
As depicted in Fig. \ref{intro}(b), these six vectors and their 2D contours represent all spin-conserving scattering events between four bands crossing \ef: the surface bands S1 and S2, and the projected bulk bands B1 and B2 \cite{pascual2006}. We observe no scattering within the same band. As shown in previous studies \cite{petersen2000,pascual2004,pascual2011} and discussed in more detail in SM Sec. \SNtheo, intra-band scattering is forbidden by spin conservation, preventing measurable real-space DoS oscillations.  

\begin{figure*}[t]
  \includegraphics[width=0.98\textwidth]{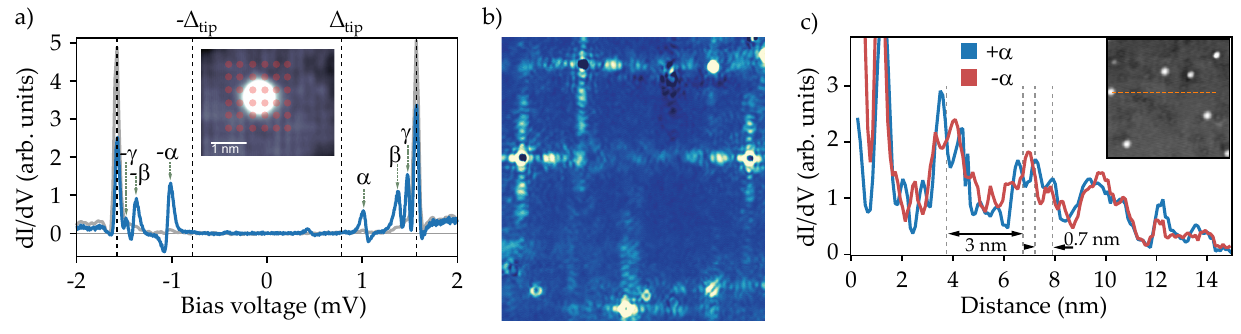}
  \caption{\textbf{Anisotropic YSR oscillations:}
  a) Spectrum on top of the V adatom (blue) shown in the topographic image in the inset (red dots represent the position of the Bi atoms on the surface). The adatom is 120~pm high. The grey curve is the spectrum of the bare \bipd\ surface. The superconducting gap of the tip, equal to the \bipd\ gap, is marked with dashed lines.  Set-point: V=-3 mV, I=300 pA, V$_{RMS}$=25 $\mu$V).
  b) Intra-gap dI/dV maps at the bias of the $\alpha_{+}$ peak (1 meV). The white color represents maximum differential conductance signal. The dark blue is the minimum. The  YSR amplitude is focused in extended beams along the (100) and (010) directions, in agreement with the nesting vectors of the surface bands. The subgap dI/dV image was measured along a previously recorded tip profile with set point V=-3 mV, I=600 pA. V$_{RMS}$=50 $\mu$V; image size: 30nm$\times$30nm.  c) Distance dependence of the particle ($\alpha_{+}$, red) and hole ($\alpha_{-}$, blue) YSR excitation's amplitude along the line in the (100) direction, shown in the topographic inset figure. The QP interference pattern shows two oscillations, agreeing with the presence of multiple bands. The extension of all YSR peaks is shown in SM Fig.~\SFQPI1.
  } \label{YSRtails}
\end{figure*}

The complex quasiparticle scattering in the normal state anticipates striking effects in the superconducting state, i.e., at subgap energies. To study these, we deposited vanadium atoms on the cold \bipd\ substrate   [inset of Fig.~\ref{YSRtails}a].  The V adatoms act as a spin-dependent scattering potential for Bogoliubov quasiparticles (BQPs), resulting in Yu-Shiba-Rusinov (YSR) bound states \cite{Yu,Shiba,Rusinov}. YSR states are localized around the impurity and are measured as subgap excitations by tunneling electrons or holes, appearing in tunneling spectra as pairs of narrow peaks inside the superconducting gap \cite{Ji2008,Heinrich2018}. High-resolution $dI/dV$ spectra measured over the V adatoms using a $\beta$-Bi$_2$Pd-coated tip \cite{Choi2018,Trivini2024} for enhancing the energy resolution \cite{Pan1998,Suderow2002,Rodrigo2004,Franke2011} shows three pairs of YSR peaks,  labeled $\alpha_{\pm}$, $\beta_{\pm}$ and $\gamma_{\pm}$ in  Fig.~\ref{YSRtails}a.  The peaks correspond to the particle (and hole, at negative bias) excitation of three YSR states formed by the interaction of three spin-polarized $d$ orbitals of V with the \bipd\ substrate. As shown in Supplementary Fig.~\SFOrbital,  the distribution of YSR signal over the V adatoms follows a pattern of lobes and nodal planes that resembles the amplitude of three different $3d$-orbitals, in accordance with the survival of a $3d_3$ valence state of V on the surface \cite{Ruby2016,Choi2017}.

\textit{Anisotropic spatial extension of YSR states:} \rev{The subgap YSR states are not simply localized around the impurity, but their amplitude extends for more than 14 nm away from the impurity with a characteristic cross-like tail along the (100) and (010) directions of the substrate [Fig.~\ref{YSRtails}b]. }
The length of these YSR tails is comparable to the coherence length of \bipd\  ($\xi_{ab}\sim$
23~nm \cite{Kacmarcik2016a,pristas2018,Imai2012,Herrera2015,Che2016}).  Such long-range patterns reflect that YSR wavefunctions far from the impurity depend on the Fermi Contour of the superconducting band hybridized with the orbital state.  The extension of BQPI patterns increases for low-dimensional bands \cite{Menard2015,Kim2020}, but also can be extended by focusing BQPs along one direction, an effect occurring for Fermi band contours with flat portions \cite{Ruby2016,ortuzar2022,uldemolins2022}. This is the case in  \bipd, where all bands around \ef\ are quasi-two dimensional and squared-shaped \cite{Sakano2015}, with abundant nesting vectors that focus the BQPI scattering along the high symmetry directions of the surface. This causes the YSR cross-shaped patterns seen in the experimental dI/dV images [Fig~\ref{YSRtails}b] and explains the long, quasi-1D decay of the YSR state \cite{ortuzar2022,uldemolins2022}.

\begin{figure*}[t]
  \includegraphics[width=0.99\textwidth]{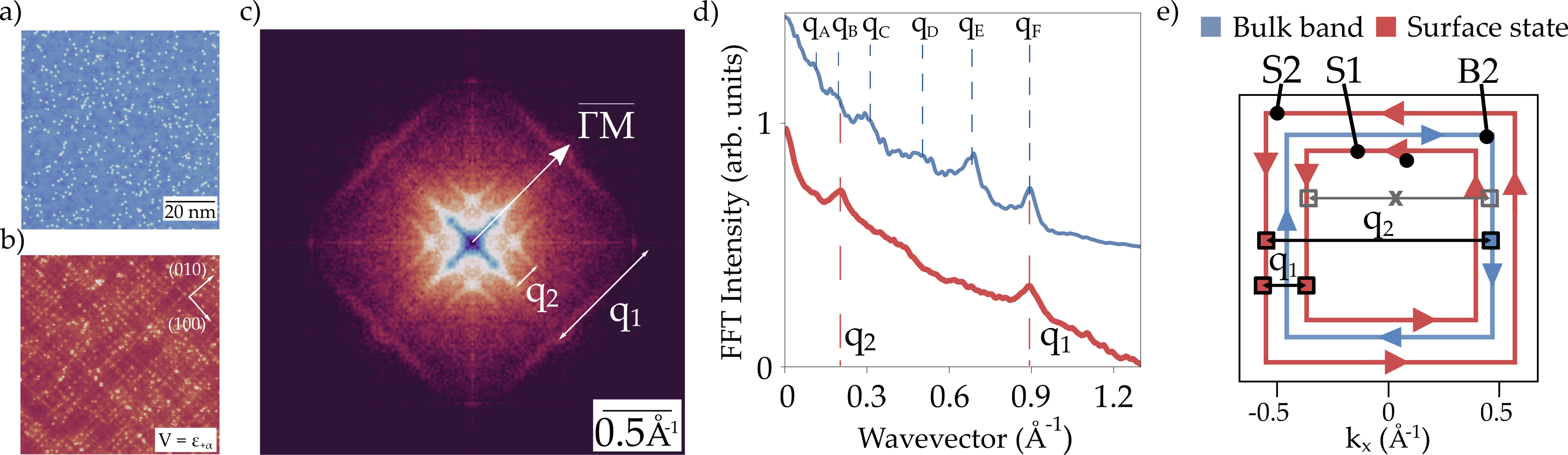}
  \caption{\textbf{Fourier transform of Bogoliubov Quasiparticle Interference:} a) Constant height current image and b) dI/dV map of the \bipd\ substrate with 1\% coverage of V adatoms.  c) Two-dimensional Fast Fourier Transform (2DFFT) of the image of panel b). The  2DFFT image shows only two contours, in contrast with the 2DFFT of the same sample at energies outside the superconducting gap shown in Fig.~\ref{intro}d.  d) Comparison of a FFT amplitude along the $\overline{\Gamma \mbox{M}}$ direction between the normal state (blue line see Fig.~\ref{intro}d) and ingap energy (red line, see white arrow in c) e) Schematic band's contours from refs. \cite{Sakano2015,Iwaya2017}, with spin helicity and allowed scattering vectors between the three bands participating in BQPI. } \label{BQPI}
\end{figure*}

\textit{Two spatial modulations:} A closer look at the YSR beams shows that their amplitude spatially  oscillates with two different wavelengths, amounting to $\lambda_1\sim$ 0.7~nm, and $\lambda_2\sim$ 3~nm. The three YSR states, $\alpha$, $\beta$ and $\gamma$, show similar long-range patterns for both particle- and hole-like excitations (SM Fig.~\SFQPI1). Figure~\ref{YSRtails}c compares particle and hole components of the $\pm \alpha$ YSR state. The shortest oscillation shows a finite particle ($p$) and hole ($h$) components dephasing. For the longest oscillation, the $p-h$ dephasing is negligible. The different dephasings can be due to a distinct scattering potential to each band.

To identify the bands involved in the YSR oscillations, we increased the coverage of V adatoms and mapped the resulting dense network of YSR oscillations for its FFT analysis. Figure \ref{BQPI}b presents an intra-gap dI/dV map of an 80 × 80 nm region of \bipd\ with 1$\%$ V atom coverage, corresponding to the topographic image in Fig.~\ref{BQPI}a. 
Despite the relatively low coverage, YSR resonances from the \alfap\ state extend from the adatoms, covering approximately 50$\%$ of the sample surface.
The FFT analysis of the dI/dV map in Fig.\ref{BQPI}b reveals two square contours representing the distribution of BQP scattering vectors \textbf{q}$_i$ in the $\alpha$ state [Fig.\ref{BQPI}c]. From these contours, we determine scattering vectors \textbf{q}$_1$ = 2.2 nm$^{-1}$ and \textbf{q}$_2$ = 8.9 nm$^{-1}$ [Fig.\ref{BQPI}d], which correspond to real-space dI/dV modulations of 3.1 nm and 0.7 nm, respectively [Fig.\ref{YSRtails}c]. 

\begin{figure}[bh!]
  \includegraphics[width=0.5\textwidth]{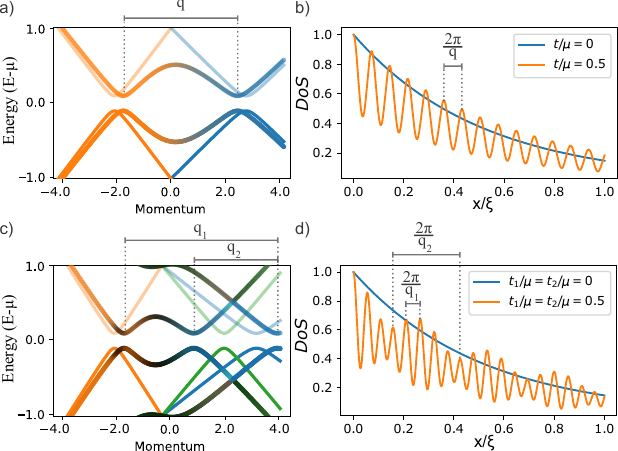}
  \caption{\textbf{Theoretical calculations:} a) and c) 1D band structure for a two-band (three-band) model. We plot the original bands (clear colored straight lines) and the hybridized ones (dark colors). b) and d) Decay of the DoS at the YSR energy for the two-band (three-band) model without (blue) and with (orange) interband hopping. The position is normalized by the coherence length $\xi$. The vectors corresponding to the observed oscillations are marked as $q_1$ and $q_2$. $t_1$ and $t_2$ refer to the hopping between first and second, and first and third bands, respectively. }\label{theo}
\end{figure}

While the shape of the FFT contours is consistent with the squared symmetry of the \bipd\ surface electronic bands \cite{ortuzar2022}, the detection of only two square features contrasts with the six scattering contours observed for normal-state QPI in Fig.~\ref{intro}c. 
In Fig.~\ref{BQPI}d, we compare the FFT intensity along $k_x$ or $\overline{\Gamma \mbox{M}}$ direction of the BQPI patterns with similar cuts in the (normal state)  QPI-FFT map of Fig.~\ref{intro}c (see SM Sec.~\SNQPI). 
The two BQPI vectors  $\textbf{q}_1$ and $\textbf{q}_2$ coincide with \textbf{q$_{\text{F}}$} and \textbf{q$_{\text{B}}$} of the (normal state) QPI pattern, respectively, while all the other scattering vectors in the normal case are missing in the YSR map. 
The scattering vectors \textbf{q$_{\text{B}}$} and \textbf{q$_{\text{F}}$} connect the outer surface band $S_2$ with the edge of the projected bulk band $\beta$ (with opposite spin helicity) and with the surface band $S1$ (same spin helicity), respectively, as shown in Fig. \ref{intro}b. Therefore, these three bands necessarily are involved in the YSR state.

\textit{Band hybridization:} Model calculations presented in SM-\SNtheo\ demonstrate that band hybridization is essential to explain the observed BQPI patterns. 
For a superconducting system with a magnetic impurity coupled to one of two independent helical bands [straight bands in Fig.~\ref{theo}a], the DoS decays smoothly with distance [blue line in Fig.~\ref{theo}b]. 
The absence of oscillations indicates that both intra-band and inter-band BQPI are forbidden, as they involve either different spin states or orthogonal bands. 
Thus, the lack of oscillations associated with intra-band scattering in the experiment agrees with the helical-like polarization of the band structure around \ef. At the same time, \rev{we conclude that the observation of two-component YSR oscillations with inter-band scattering vectors necessarily implies that the participating bands are hybridized.}
 
To account for hybridization between bands, we inserted inter-band hopping elements in the model of SM-\SNtheo, which coupled the two bands, as shown in Fig.\ref{theo}a. The newly formed hybrid bands 
now allow for the emergence of BQPI oscillations in the DoS   [Fig.\ref{theo}b]. 
Consequently, band hybridization is crucial for generating BQPI (as well as in normal-state QPI) from a single impurity in a system with orthogonal helical bands. It is worth noting that the strong band mixing in the superconducting state of \bipd\ is fully consistent with its single-gap superconductivity. Furthermore, this behavior can be attributed to the significant band overlap in momentum space, as revealed by DFT calculations \cite{Sakano2015,Iwaya2017}.

Intriguingly, the BQPI pattern exhibits only \rev{two scattering vectors, \textbf{q}$_{\text{1}}=$\textbf{q}$_{\text{F}}$ and \textbf{q}$_{\text{2}}=$\textbf{q}$_{\text{B}}$, connecting the outer band $S2$  with bands $S1$ and $B2$  [Fig. \ref{BQPI}e], while the third expected vector \textbf{q}$_{\text{E}}$ connecting $S1$ with $B2$ is absent. 
This selectivity arises because YSR excitations, being sub-gap states, are restricted to bands hybridized with the magnetic impurity. 
The predominant role of S2 in the  BQPI patterns indicates that this band forms the YSR channel through its hybridization with the vanadium $d$ orbitals.  Consequently, only hybrid pairs with spectral weight in S2 contribute to the BQPI at sub-gap energies, while other hybrid pairs, such as those mixing bands S1 and B2 (connected through \textbf{q}$_E$), do not couple with the impurity, and their quasiparticle excitation energy remains outside the bulk gap.}

To demonstrate such selective band interference, we simulate in Fig.~\ref{theo}c the DoS of three helical bands, with only one of them coupled to a magnetic impurity. In the absence of interband hopping, the DoS decays smoothly away from the impurity without oscillations  [blue line in Fig.~\ref{theo}d]. Mixing all the bands with interband hopping elements leads to a single hybrid band with three identical gaps, from which scattering vectors can be obtained. In real space, this configuration results in a two-component BQPI pattern similar to the one we observe in the experiment [orange line in Fig.~\ref{theo}d]. 
Thus, the experimental patterns reflect band mixing effects in the superconducting condensate through the interference pattern of their \textit{fermionic} excitations.  

Above the superconducting gap, all bands are degenerate, and their quasiparticles are sensitive to potential scattering by defects or impurities. In this scenario, all spin-conserving QPI emerges either because interband hopping mixes all bands or, simply, by multi-impurity scattering (both mechanisms are effectively similar, as shown in the SM-\SNtheoo). 


\textit{Discussion:} The multi-frequency interference patterns of \bipd\ provide interesting 
information on pairing mechanisms in this material. First, the lack of BQPI patterns associated with S2 intra-band scattering corroborates its predicted helical nature. Consequently, the pairing of electrons within this band involves a mixture of singlet and triplet components because an intraband singlet cannot be formed (as depicted in Fig.~\ref{intro0}) .  
Furthermore, exotic pairing schemes such as triplet or odd-frequency are expected from the hybridization of spin-helical bands with different orbital symmetries  
\cite{Black-Schaffer2013,Bergeret2005,bergeret2023triplet,kanasugi2022}.  

Since the mixed bands also have different wave vectors, the \textit{hybrid} Cooper pairs should have a finite momentum. This case resembles the Fulde–Ferrell–Larkin–Ovchinnikov (FFLO) states \cite{Fulde1964,larkin1965inhomogeneous} but produced with helical states instead of the Zeeman-split Fermi surface \cite{Asaba2024}.  In this case, the momentum averages to zero, but an in-plane magnetic field can induce a net Cooper pair momentum,  leading to non-reciprocal currents \cite{Yuan2022,ilic2022theory}.
Although experimental signatures of such unconventional superconducting phases are not directly evident from the BQPI patterns, our measurements reveal key ingredients for detecting their existence. \rev{We foresee that this kind of BQPI studies, in combination with in-plane magnetic fields \cite{Powell2025}, could  provide further evidence of exotic pairing schemes enabled by multi-band coupling.}



\begin{acknowledgments}
We acknowledge financial support from the Spanish MCIN/AEI/10.13039/501100011033 and the European Regional Development Fund (ERDF) through grants 
PID2020-114071RB-I00, 
PID2020-114252GB-I00, 
PID2022-140845OB-C61, 
PID2023-148225NB-C31, 
PID2023-150148OB-I00, 
CEX2020-001038-M, 
CEX2023-001316-M, 
TED2021-130292B-C42, 
and TED2021-130546B-I00, 
from the Basque Government through grant IT-1591-22, 
from the European Union NextGenerationEU/PRTR-C17.I1 through the IKUR Strategy of the Department of Education of the Basque Government under a collaboration agreement with Ikerbasque MPC, CIC nanoGUNE, and DIPC, 
and from the European Union’s Horizon Europe through grant JOSEPHINE (No. 101130224) and through ERC-AdG CONSPIRA (No. 101097693). H.S. and E.H. acknowledge the QUASURF project [ref. SI4/PJI/2024-00062] funded by the Comunidad de Madrid through the agreement to promote and encourage research and technology transfer at the Universidad Autónoma de Madrid.
We also acknowledge collaborations through EU program Cost CA21144 (Superqumap).  J.O. acknowledges the scholarship PRE-2021-1-0350 from the Basque Government.

\end{acknowledgments}


\begin{thebibliography}{45}%
\makeatletter
\providecommand \@ifxundefined [1]{%
 \@ifx{#1\undefined}
}%
\providecommand \@ifnum [1]{%
 \ifnum #1\expandafter \@firstoftwo
 \else \expandafter \@secondoftwo
 \fi
}%
\providecommand \@ifx [1]{%
 \ifx #1\expandafter \@firstoftwo
 \else \expandafter \@secondoftwo
 \fi
}%
\providecommand \natexlab [1]{#1}%
\providecommand \enquote  [1]{``#1''}%
\providecommand \bibnamefont  [1]{#1}%
\providecommand \bibfnamefont [1]{#1}%
\providecommand \citenamefont [1]{#1}%
\providecommand \href@noop [0]{\@secondoftwo}%
\providecommand \href [0]{\begingroup \@sanitize@url \@href}%
\providecommand \@href[1]{\@@startlink{#1}\@@href}%
\providecommand \@@href[1]{\endgroup#1\@@endlink}%
\providecommand \@sanitize@url [0]{\catcode `\\12\catcode `\$12\catcode `\&12\catcode `\#12\catcode `\^12\catcode `\_12\catcode `\%12\relax}%
\providecommand \@@startlink[1]{}%
\providecommand \@@endlink[0]{}%
\providecommand \url  [0]{\begingroup\@sanitize@url \@url }%
\providecommand \@url [1]{\endgroup\@href {#1}{\urlprefix }}%
\providecommand \urlprefix  [0]{URL }%
\providecommand \Eprint [0]{\href }%
\providecommand \doibase [0]{http://dx.doi.org/}%
\providecommand \selectlanguage [0]{\@gobble}%
\providecommand \bibinfo  [0]{\@secondoftwo}%
\providecommand \bibfield  [0]{\@secondoftwo}%
\providecommand \translation [1]{[#1]}%
\providecommand \BibitemOpen [0]{}%
\providecommand \bibitemStop [0]{}%
\providecommand \bibitemNoStop [0]{.\EOS\space}%
\providecommand \EOS [0]{\spacefactor3000\relax}%
\providecommand \BibitemShut  [1]{\csname bibitem#1\endcsname}%
\let\auto@bib@innerbib\@empty
\bibitem [{\citenamefont {Black-Schaffer}\ and\ \citenamefont {Balatsky}(2013)}]{Black-Schaffer2013}%
  \BibitemOpen
  \bibfield  {author} {\bibinfo {author} {\bibfnamefont {A.~M.}\ \bibnamefont {Black-Schaffer}}\ and\ \bibinfo {author} {\bibfnamefont {A.~V.}\ \bibnamefont {Balatsky}},\ }\href {\doibase 10.1103/PhysRevB.88.104514} {\bibfield  {journal} {\bibinfo  {journal} {Phys. Rev. B - Condens. Matter Mater. Phys.}\ }\textbf {\bibinfo {volume} {88}},\ \bibinfo {pages} {1} (\bibinfo {year} {2013})}\BibitemShut {NoStop}%
\bibitem [{\citenamefont {Gor'kov}\ and\ \citenamefont {Rashba}(2001)}]{Gorkov2001}%
  \BibitemOpen
  \bibfield  {author} {\bibinfo {author} {\bibfnamefont {L.~P.}\ \bibnamefont {Gor'kov}}\ and\ \bibinfo {author} {\bibfnamefont {E.~I.}\ \bibnamefont {Rashba}},\ }\href {\doibase 10.1103/PhysRevLett.87.037004} {\bibfield  {journal} {\bibinfo  {journal} {Phys. Rev. Lett.}\ }\textbf {\bibinfo {volume} {87}},\ \bibinfo {pages} {37004} (\bibinfo {year} {2001})},\ \Eprint {http://arxiv.org/abs/0103449} {arXiv:0103449 [cond-mat]} \BibitemShut {NoStop}%
\bibitem [{\citenamefont {Kim}\ \emph {et~al.}(2015)\citenamefont {Kim}, \citenamefont {Zhang}, \citenamefont {Rossi},\ and\ \citenamefont {Lutchyn}}]{Kim2015}%
  \BibitemOpen
  \bibfield  {author} {\bibinfo {author} {\bibfnamefont {Y.}~\bibnamefont {Kim}}, \bibinfo {author} {\bibfnamefont {J.}~\bibnamefont {Zhang}}, \bibinfo {author} {\bibfnamefont {E.}~\bibnamefont {Rossi}}, \ and\ \bibinfo {author} {\bibfnamefont {R.~M.}\ \bibnamefont {Lutchyn}},\ }\href {\doibase 10.1103/PhysRevLett.114.236804} {\bibfield  {journal} {\bibinfo  {journal} {Phys. Rev. Lett.}\ }\textbf {\bibinfo {volume} {114}},\ \bibinfo {pages} {236804} (\bibinfo {year} {2015})},\ \Eprint {http://arxiv.org/abs/arXiv:1410.4558v3} {arXiv:arXiv:1410.4558v3} \BibitemShut {NoStop}%
\bibitem [{\citenamefont {Sato}\ and\ \citenamefont {Ando}(2017)}]{Sato2017}%
  \BibitemOpen
  \bibfield  {author} {\bibinfo {author} {\bibfnamefont {M.}~\bibnamefont {Sato}}\ and\ \bibinfo {author} {\bibfnamefont {Y.}~\bibnamefont {Ando}},\ }\href {\doibase 10.1088/1361-6633/aa6ac7} {\bibfield  {journal} {\bibinfo  {journal} {Reports on Progress in Physics}\ }\textbf {\bibinfo {volume} {80}},\ \bibinfo {pages} {076501} (\bibinfo {year} {2017})}\BibitemShut {NoStop}%
\bibitem [{\citenamefont {Asaba}\ \emph {et~al.}(2024)\citenamefont {Asaba}, \citenamefont {Naritsuka}, \citenamefont {Asaeda}, \citenamefont {Kosuge}, \citenamefont {Ikemori}, \citenamefont {Suetsugu}, \citenamefont {Kasahara}, \citenamefont {Kohsaka}, \citenamefont {Terashima}, \citenamefont {Daido}, \citenamefont {Yanase},\ and\ \citenamefont {Matsuda}}]{Asaba2024}%
  \BibitemOpen
  \bibfield  {author} {\bibinfo {author} {\bibfnamefont {T.}~\bibnamefont {Asaba}}, \bibinfo {author} {\bibfnamefont {M.}~\bibnamefont {Naritsuka}}, \bibinfo {author} {\bibfnamefont {H.}~\bibnamefont {Asaeda}}, \bibinfo {author} {\bibfnamefont {Y.}~\bibnamefont {Kosuge}}, \bibinfo {author} {\bibfnamefont {S.}~\bibnamefont {Ikemori}}, \bibinfo {author} {\bibfnamefont {S.}~\bibnamefont {Suetsugu}}, \bibinfo {author} {\bibfnamefont {Y.}~\bibnamefont {Kasahara}}, \bibinfo {author} {\bibfnamefont {Y.}~\bibnamefont {Kohsaka}}, \bibinfo {author} {\bibfnamefont {T.}~\bibnamefont {Terashima}}, \bibinfo {author} {\bibfnamefont {A.}~\bibnamefont {Daido}}, \bibinfo {author} {\bibfnamefont {Y.}~\bibnamefont {Yanase}}, \ and\ \bibinfo {author} {\bibfnamefont {Y.}~\bibnamefont {Matsuda}},\ }\href {\doibase 10.1038/s41467-024-47875-4} {\bibfield  {journal} {\bibinfo  {journal} {Nature Communications}\ }\textbf {\bibinfo {volume} {15}},\ \bibinfo {pages} {3861} (\bibinfo {year} {2024})}\BibitemShut {NoStop}%
\bibitem [{\citenamefont {Sprau}\ \emph {et~al.}(2017)\citenamefont {Sprau}, \citenamefont {Kostin}, \citenamefont {Kreisel}, \citenamefont {B{\"{o}}hmer}, \citenamefont {Taufour}, \citenamefont {Canfield}, \citenamefont {Mukherjee}, \citenamefont {Hirschfeld}, \citenamefont {Andersen},\ and\ \citenamefont {Davis}}]{Sprau2017}%
  \BibitemOpen
  \bibfield  {author} {\bibinfo {author} {\bibfnamefont {P.~O.}\ \bibnamefont {Sprau}}, \bibinfo {author} {\bibfnamefont {A.}~\bibnamefont {Kostin}}, \bibinfo {author} {\bibfnamefont {A.}~\bibnamefont {Kreisel}}, \bibinfo {author} {\bibfnamefont {A.~E.}\ \bibnamefont {B{\"{o}}hmer}}, \bibinfo {author} {\bibfnamefont {V.}~\bibnamefont {Taufour}}, \bibinfo {author} {\bibfnamefont {P.~C.}\ \bibnamefont {Canfield}}, \bibinfo {author} {\bibfnamefont {S.}~\bibnamefont {Mukherjee}}, \bibinfo {author} {\bibfnamefont {P.~J.}\ \bibnamefont {Hirschfeld}}, \bibinfo {author} {\bibfnamefont {B.~M.}\ \bibnamefont {Andersen}}, \ and\ \bibinfo {author} {\bibfnamefont {J.~C.}\ \bibnamefont {Davis}},\ }\href {\doibase 10.1126/science.aal1575} {\bibfield  {journal} {\bibinfo  {journal} {Science}\ }\textbf {\bibinfo {volume} {357}},\ \bibinfo {pages} {75} (\bibinfo {year} {2017})},\ \Eprint {http://arxiv.org/abs/1611.02134} {arXiv:1611.02134} \BibitemShut {NoStop}%
\bibitem [{\citenamefont {Sharma}\ \emph {et~al.}(2020)\citenamefont {Sharma}, \citenamefont {Edkins}, \citenamefont {Wang}, \citenamefont {Kostin}, \citenamefont {Sow}, \citenamefont {Maeno}, \citenamefont {Mackenzie}, \citenamefont {{S{\'{e}}amus Davis}},\ and\ \citenamefont {Madhavan}}]{Sharma2020}%
  \BibitemOpen
  \bibfield  {author} {\bibinfo {author} {\bibfnamefont {R.}~\bibnamefont {Sharma}}, \bibinfo {author} {\bibfnamefont {S.~D.}\ \bibnamefont {Edkins}}, \bibinfo {author} {\bibfnamefont {Z.}~\bibnamefont {Wang}}, \bibinfo {author} {\bibfnamefont {A.}~\bibnamefont {Kostin}}, \bibinfo {author} {\bibfnamefont {C.}~\bibnamefont {Sow}}, \bibinfo {author} {\bibfnamefont {Y.}~\bibnamefont {Maeno}}, \bibinfo {author} {\bibfnamefont {A.~P.}\ \bibnamefont {Mackenzie}}, \bibinfo {author} {\bibfnamefont {J.~C.}\ \bibnamefont {{S{\'{e}}amus Davis}}}, \ and\ \bibinfo {author} {\bibfnamefont {V.}~\bibnamefont {Madhavan}},\ }\href {\doibase 10.1073/pnas.1916463117} {\bibfield  {journal} {\bibinfo  {journal} {Proc. Natl. Acad. Sci. U. S. A.}\ }\textbf {\bibinfo {volume} {117}},\ \bibinfo {pages} {5222} (\bibinfo {year} {2020})}\BibitemShut {NoStop}%
\bibitem [{\citenamefont {Sakano}\ \emph {et~al.}(2015)\citenamefont {Sakano}, \citenamefont {Okawa}, \citenamefont {Kanou}, \citenamefont {Sanjo}, \citenamefont {Okuda}, \citenamefont {Sasagawa},\ and\ \citenamefont {Ishizaka}}]{Sakano2015}%
  \BibitemOpen
  \bibfield  {author} {\bibinfo {author} {\bibfnamefont {M.}~\bibnamefont {Sakano}}, \bibinfo {author} {\bibfnamefont {K.}~\bibnamefont {Okawa}}, \bibinfo {author} {\bibfnamefont {M.}~\bibnamefont {Kanou}}, \bibinfo {author} {\bibfnamefont {H.}~\bibnamefont {Sanjo}}, \bibinfo {author} {\bibfnamefont {T.}~\bibnamefont {Okuda}}, \bibinfo {author} {\bibfnamefont {T.}~\bibnamefont {Sasagawa}}, \ and\ \bibinfo {author} {\bibfnamefont {K.}~\bibnamefont {Ishizaka}},\ }\href {\doibase 10.1038/ncomms9595} {\bibfield  {journal} {\bibinfo  {journal} {Nat. Commun.}\ }\textbf {\bibinfo {volume} {6}},\ \bibinfo {pages} {8595} (\bibinfo {year} {2015})},\ \Eprint {http://arxiv.org/abs/1505.07231} {arXiv:1505.07231} \BibitemShut {NoStop}%
\bibitem [{\citenamefont {Iwaya}\ \emph {et~al.}(2017)\citenamefont {Iwaya}, \citenamefont {Kohsaka}, \citenamefont {Okawa}, \citenamefont {Machida}, \citenamefont {Bahramy}, \citenamefont {Hanaguri},\ and\ \citenamefont {Sasagawa}}]{Iwaya2017}%
  \BibitemOpen
  \bibfield  {author} {\bibinfo {author} {\bibfnamefont {K.}~\bibnamefont {Iwaya}}, \bibinfo {author} {\bibfnamefont {Y.}~\bibnamefont {Kohsaka}}, \bibinfo {author} {\bibfnamefont {K.}~\bibnamefont {Okawa}}, \bibinfo {author} {\bibfnamefont {T.}~\bibnamefont {Machida}}, \bibinfo {author} {\bibfnamefont {M.~S.}\ \bibnamefont {Bahramy}}, \bibinfo {author} {\bibfnamefont {T.}~\bibnamefont {Hanaguri}}, \ and\ \bibinfo {author} {\bibfnamefont {T.}~\bibnamefont {Sasagawa}},\ }\href {\doibase 10.1038/s41467-017-01209-9} {\bibfield  {journal} {\bibinfo  {journal} {Nat. Commun.}\ }\textbf {\bibinfo {volume} {8}},\ \bibinfo {pages} {976} (\bibinfo {year} {2017})}\BibitemShut {NoStop}%
\bibitem [{\citenamefont {Petersen}\ and\ \citenamefont {Hedeg{\aa}rd}(2000)}]{petersen2000}%
  \BibitemOpen
  \bibfield  {author} {\bibinfo {author} {\bibfnamefont {L.}~\bibnamefont {Petersen}}\ and\ \bibinfo {author} {\bibfnamefont {P.}~\bibnamefont {Hedeg{\aa}rd}},\ }\href {\doibase 10.1016/S0039-6028(00)00441-6} {\bibfield  {journal} {\bibinfo  {journal} {Surface Science}\ }\textbf {\bibinfo {volume} {459}},\ \bibinfo {pages} {49} (\bibinfo {year} {2000})}\BibitemShut {NoStop}%
\bibitem [{\citenamefont {Pascual}\ \emph {et~al.}(2004)\citenamefont {Pascual}, \citenamefont {Bihlmayer}, \citenamefont {Koroteev}, \citenamefont {Rust}, \citenamefont {Ceballos}, \citenamefont {Hansmann}, \citenamefont {Horn}, \citenamefont {Chulkov}, \citenamefont {Bl{\"u}gel}, \citenamefont {Echenique},\ and\ \citenamefont {Hofmann}}]{pascual2004}%
  \BibitemOpen
  \bibfield  {author} {\bibinfo {author} {\bibfnamefont {J.~I.}\ \bibnamefont {Pascual}}, \bibinfo {author} {\bibfnamefont {G.}~\bibnamefont {Bihlmayer}}, \bibinfo {author} {\bibfnamefont {{\relax Yu}.~M.}\ \bibnamefont {Koroteev}}, \bibinfo {author} {\bibfnamefont {H.-P.}\ \bibnamefont {Rust}}, \bibinfo {author} {\bibfnamefont {G.}~\bibnamefont {Ceballos}}, \bibinfo {author} {\bibfnamefont {M.}~\bibnamefont {Hansmann}}, \bibinfo {author} {\bibfnamefont {K.}~\bibnamefont {Horn}}, \bibinfo {author} {\bibfnamefont {E.~V.}\ \bibnamefont {Chulkov}}, \bibinfo {author} {\bibfnamefont {S.}~\bibnamefont {Bl{\"u}gel}}, \bibinfo {author} {\bibfnamefont {P.~M.}\ \bibnamefont {Echenique}}, \ and\ \bibinfo {author} {\bibfnamefont {{\relax Ph}.}~\bibnamefont {Hofmann}},\ }\href {\doibase 10.1103/PhysRevLett.93.196802} {\bibfield  {journal} {\bibinfo  {journal} {Physical Review Letters}\ }\textbf {\bibinfo {volume} {93}},\ \bibinfo {pages} {196802} (\bibinfo {year} {2004})}\BibitemShut {NoStop}%
\bibitem [{\citenamefont {Str\'o\ifmmode~\dot{z}\else \.{z}\fi{}ecka}\ \emph {et~al.}(2011)\citenamefont {Str\'o\ifmmode~\dot{z}\else \.{z}\fi{}ecka}, \citenamefont {Eiguren},\ and\ \citenamefont {Pascual}}]{pascual2011}%
  \BibitemOpen
  \bibfield  {author} {\bibinfo {author} {\bibfnamefont {A.}~\bibnamefont {Str\'o\ifmmode~\dot{z}\else \.{z}\fi{}ecka}}, \bibinfo {author} {\bibfnamefont {A.}~\bibnamefont {Eiguren}}, \ and\ \bibinfo {author} {\bibfnamefont {J.~I.}\ \bibnamefont {Pascual}},\ }\href {\doibase 10.1103/PhysRevLett.107.186805} {\bibfield  {journal} {\bibinfo  {journal} {Phys. Rev. Lett.}\ }\textbf {\bibinfo {volume} {107}},\ \bibinfo {pages} {186805} (\bibinfo {year} {2011})}\BibitemShut {NoStop}%
\bibitem [{\citenamefont {Steinbrecher}\ \emph {et~al.}(2013)\citenamefont {Steinbrecher}, \citenamefont {Harutyunyan}, \citenamefont {Ast},\ and\ \citenamefont {Wegner}}]{Steinbrecher2013}%
  \BibitemOpen
  \bibfield  {author} {\bibinfo {author} {\bibfnamefont {M.}~\bibnamefont {Steinbrecher}}, \bibinfo {author} {\bibfnamefont {H.}~\bibnamefont {Harutyunyan}}, \bibinfo {author} {\bibfnamefont {C.~R.}\ \bibnamefont {Ast}}, \ and\ \bibinfo {author} {\bibfnamefont {D.}~\bibnamefont {Wegner}},\ }\href {\doibase 10.1103/PhysRevB.87.245436} {\bibfield  {journal} {\bibinfo  {journal} {Phys. Rev. B}\ }\textbf {\bibinfo {volume} {87}},\ \bibinfo {pages} {245436} (\bibinfo {year} {2013})}\BibitemShut {NoStop}%
\bibitem [{\citenamefont {Herrera}\ \emph {et~al.}(2023)\citenamefont {Herrera}, \citenamefont {Guillam{\'o}n}, \citenamefont {Barrena}, \citenamefont {Herrera}, \citenamefont {Galvis}, \citenamefont {Yeyati}, \citenamefont {Rusz}, \citenamefont {Oppeneer}, \citenamefont {Knebel}, \citenamefont {Brison}, \citenamefont {Flouquet}, \citenamefont {Aoki},\ and\ \citenamefont {Suderow}}]{Herrera2023}%
  \BibitemOpen
  \bibfield  {author} {\bibinfo {author} {\bibfnamefont {E.}~\bibnamefont {Herrera}}, \bibinfo {author} {\bibfnamefont {I.}~\bibnamefont {Guillam{\'o}n}}, \bibinfo {author} {\bibfnamefont {V.}~\bibnamefont {Barrena}}, \bibinfo {author} {\bibfnamefont {W.~J.}\ \bibnamefont {Herrera}}, \bibinfo {author} {\bibfnamefont {J.~A.}\ \bibnamefont {Galvis}}, \bibinfo {author} {\bibfnamefont {A.~L.}\ \bibnamefont {Yeyati}}, \bibinfo {author} {\bibfnamefont {J.}~\bibnamefont {Rusz}}, \bibinfo {author} {\bibfnamefont {P.~M.}\ \bibnamefont {Oppeneer}}, \bibinfo {author} {\bibfnamefont {G.}~\bibnamefont {Knebel}}, \bibinfo {author} {\bibfnamefont {J.~P.}\ \bibnamefont {Brison}}, \bibinfo {author} {\bibfnamefont {J.}~\bibnamefont {Flouquet}}, \bibinfo {author} {\bibfnamefont {D.}~\bibnamefont {Aoki}}, \ and\ \bibinfo {author} {\bibfnamefont {H.}~\bibnamefont {Suderow}},\ }\href {\doibase 10.1038/s41586-023-05830-1} {\bibfield  {journal} {\bibinfo  {journal} {Nature}\ }\textbf {\bibinfo {volume} {616}},\ \bibinfo {pages}
  {465} (\bibinfo {year} {2023})}\BibitemShut {NoStop}%
\bibitem [{\citenamefont {Imai}\ \emph {et~al.}(2012)\citenamefont {Imai}, \citenamefont {Nabeshima}, \citenamefont {Yoshinaka}, \citenamefont {Miyatani}, \citenamefont {Kondo}, \citenamefont {Komiya}, \citenamefont {Tsukada},\ and\ \citenamefont {Maeda}}]{Imai2012}%
  \BibitemOpen
  \bibfield  {author} {\bibinfo {author} {\bibfnamefont {Y.}~\bibnamefont {Imai}}, \bibinfo {author} {\bibfnamefont {F.}~\bibnamefont {Nabeshima}}, \bibinfo {author} {\bibfnamefont {T.}~\bibnamefont {Yoshinaka}}, \bibinfo {author} {\bibfnamefont {K.}~\bibnamefont {Miyatani}}, \bibinfo {author} {\bibfnamefont {R.}~\bibnamefont {Kondo}}, \bibinfo {author} {\bibfnamefont {S.}~\bibnamefont {Komiya}}, \bibinfo {author} {\bibfnamefont {I.}~\bibnamefont {Tsukada}}, \ and\ \bibinfo {author} {\bibfnamefont {A.}~\bibnamefont {Maeda}},\ }\href {\doibase 10.1143/JPSJ.81.113708} {\bibfield  {journal} {\bibinfo  {journal} {J. Phys. Soc. Japan}\ }\textbf {\bibinfo {volume} {81}},\ \bibinfo {pages} {113708} (\bibinfo {year} {2012})}\BibitemShut {NoStop}%
\bibitem [{\citenamefont {Herrera}\ \emph {et~al.}(2015)\citenamefont {Herrera}, \citenamefont {Guillam{\'{o}}n}, \citenamefont {Galvis}, \citenamefont {Correa}, \citenamefont {Fente}, \citenamefont {Luccas}, \citenamefont {Mompean}, \citenamefont {Garc{\'{i}}a-Hern{\'{a}}ndez}, \citenamefont {Vieira}, \citenamefont {Brison},\ and\ \citenamefont {Suderow}}]{Herrera2015}%
  \BibitemOpen
  \bibfield  {author} {\bibinfo {author} {\bibfnamefont {E.}~\bibnamefont {Herrera}}, \bibinfo {author} {\bibfnamefont {I.}~\bibnamefont {Guillam{\'{o}}n}}, \bibinfo {author} {\bibfnamefont {J.~A.}\ \bibnamefont {Galvis}}, \bibinfo {author} {\bibfnamefont {A.}~\bibnamefont {Correa}}, \bibinfo {author} {\bibfnamefont {A.}~\bibnamefont {Fente}}, \bibinfo {author} {\bibfnamefont {R.~F.}\ \bibnamefont {Luccas}}, \bibinfo {author} {\bibfnamefont {F.~J.}\ \bibnamefont {Mompean}}, \bibinfo {author} {\bibfnamefont {M.}~\bibnamefont {Garc{\'{i}}a-Hern{\'{a}}ndez}}, \bibinfo {author} {\bibfnamefont {S.}~\bibnamefont {Vieira}}, \bibinfo {author} {\bibfnamefont {J.~P.}\ \bibnamefont {Brison}}, \ and\ \bibinfo {author} {\bibfnamefont {H.}~\bibnamefont {Suderow}},\ }\href {\doibase 10.1103/PhysRevB.92.054507} {\bibfield  {journal} {\bibinfo  {journal} {Phys. Rev. B}\ }\textbf {\bibinfo {volume} {92}},\ \bibinfo {pages} {1} (\bibinfo {year} {2015})}\BibitemShut {NoStop}%
\bibitem [{\citenamefont {Pascual}\ \emph {et~al.}(2006)\citenamefont {Pascual}, \citenamefont {Dick}, \citenamefont {Hansmann}, \citenamefont {Rust}, \citenamefont {Neugebauer},\ and\ \citenamefont {Horn}}]{pascual2006}%
  \BibitemOpen
  \bibfield  {author} {\bibinfo {author} {\bibfnamefont {J.~I.}\ \bibnamefont {Pascual}}, \bibinfo {author} {\bibfnamefont {A.}~\bibnamefont {Dick}}, \bibinfo {author} {\bibfnamefont {M.}~\bibnamefont {Hansmann}}, \bibinfo {author} {\bibfnamefont {H.-P.}\ \bibnamefont {Rust}}, \bibinfo {author} {\bibfnamefont {J.}~\bibnamefont {Neugebauer}}, \ and\ \bibinfo {author} {\bibfnamefont {K.}~\bibnamefont {Horn}},\ }\href {\doibase 10.1103/PhysRevLett.96.046801} {\bibfield  {journal} {\bibinfo  {journal} {Phys. Rev. Lett.}\ }\textbf {\bibinfo {volume} {96}},\ \bibinfo {pages} {046801} (\bibinfo {year} {2006})}\BibitemShut {NoStop}%
\bibitem [{\citenamefont {Yu}(1965)}]{Yu}%
  \BibitemOpen
  \bibfield  {author} {\bibinfo {author} {\bibfnamefont {L.}~\bibnamefont {Yu}},\ }\href {\doibase 10.7498/aps.21.75} {\bibfield  {journal} {\bibinfo  {journal} {Acta Physica Sinica}\ }\textbf {\bibinfo {volume} {21}},\ \bibinfo {pages} {75} (\bibinfo {year} {1965})}\BibitemShut {NoStop}%
\bibitem [{\citenamefont {Shiba}(1968)}]{Shiba}%
  \BibitemOpen
  \bibfield  {author} {\bibinfo {author} {\bibfnamefont {H.}~\bibnamefont {Shiba}},\ }\href {\doibase 10.1143/PTP.40.435} {\bibfield  {journal} {\bibinfo  {journal} {Prog. Theor. Phys.}\ }\textbf {\bibinfo {volume} {40}},\ \bibinfo {pages} {435} (\bibinfo {year} {1968})}\BibitemShut {NoStop}%
\bibitem [{\citenamefont {{Rusinov}}(1969)}]{Rusinov}%
  \BibitemOpen
  \bibfield  {author} {\bibinfo {author} {\bibfnamefont {A.~I.}\ \bibnamefont {{Rusinov}}},\ }\href {http://adsabs.harvard.edu/abs/1969JETP...29.1101R} {\bibfield  {journal} {\bibinfo  {journal} {Sov. J. Exp. Theor. Phys.}\ }\textbf {\bibinfo {volume} {29}},\ \bibinfo {pages} {1101} (\bibinfo {year} {1969})}\BibitemShut {NoStop}%
\bibitem [{\citenamefont {Ji}\ \emph {et~al.}(2008)\citenamefont {Ji}, \citenamefont {Zhang}, \citenamefont {Fu}, \citenamefont {Chen}, \citenamefont {Ma}, \citenamefont {Li}, \citenamefont {Duan}, \citenamefont {Jia},\ and\ \citenamefont {Xue}}]{Ji2008}%
  \BibitemOpen
  \bibfield  {author} {\bibinfo {author} {\bibfnamefont {S.-H.}\ \bibnamefont {Ji}}, \bibinfo {author} {\bibfnamefont {T.}~\bibnamefont {Zhang}}, \bibinfo {author} {\bibfnamefont {Y.-S.}\ \bibnamefont {Fu}}, \bibinfo {author} {\bibfnamefont {X.}~\bibnamefont {Chen}}, \bibinfo {author} {\bibfnamefont {X.-C.}\ \bibnamefont {Ma}}, \bibinfo {author} {\bibfnamefont {J.}~\bibnamefont {Li}}, \bibinfo {author} {\bibfnamefont {W.-H.}\ \bibnamefont {Duan}}, \bibinfo {author} {\bibfnamefont {J.-F.}\ \bibnamefont {Jia}}, \ and\ \bibinfo {author} {\bibfnamefont {Q.-K.}\ \bibnamefont {Xue}},\ }\href {\doibase 10.1103/PhysRevLett.100.226801} {\bibfield  {journal} {\bibinfo  {journal} {Phys. Rev. Lett.}\ }\textbf {\bibinfo {volume} {100}},\ \bibinfo {pages} {226801} (\bibinfo {year} {2008})}\BibitemShut {NoStop}%
\bibitem [{\citenamefont {Heinrich}\ \emph {et~al.}(2018)\citenamefont {Heinrich}, \citenamefont {Pascual},\ and\ \citenamefont {Franke}}]{Heinrich2018}%
  \BibitemOpen
  \bibfield  {author} {\bibinfo {author} {\bibfnamefont {B.~W.}\ \bibnamefont {Heinrich}}, \bibinfo {author} {\bibfnamefont {J.~I.}\ \bibnamefont {Pascual}}, \ and\ \bibinfo {author} {\bibfnamefont {K.~J.}\ \bibnamefont {Franke}},\ }\href {\doibase 10.1016/j.progsurf.2018.01.001} {\bibfield  {journal} {\bibinfo  {journal} {Prog. Surf. Sci.}\ }\textbf {\bibinfo {volume} {93}},\ \bibinfo {pages} {1} (\bibinfo {year} {2018})},\ \Eprint {http://arxiv.org/abs/1705.03672} {arXiv:1705.03672} \BibitemShut {NoStop}%
\bibitem [{\citenamefont {Choi}\ \emph {et~al.}(2018)\citenamefont {Choi}, \citenamefont {Fern{\'{a}}ndez}, \citenamefont {Herrera}, \citenamefont {Rubio-Verd{\'{u}}}, \citenamefont {Ugeda}, \citenamefont {Guillam{\'{o}}n}, \citenamefont {Suderow}, \citenamefont {Pascual},\ and\ \citenamefont {Lorente}}]{Choi2018}%
  \BibitemOpen
  \bibfield  {author} {\bibinfo {author} {\bibfnamefont {D.~J.}\ \bibnamefont {Choi}}, \bibinfo {author} {\bibfnamefont {C.~G.}\ \bibnamefont {Fern{\'{a}}ndez}}, \bibinfo {author} {\bibfnamefont {E.}~\bibnamefont {Herrera}}, \bibinfo {author} {\bibfnamefont {C.}~\bibnamefont {Rubio-Verd{\'{u}}}}, \bibinfo {author} {\bibfnamefont {M.~M.}\ \bibnamefont {Ugeda}}, \bibinfo {author} {\bibfnamefont {I.}~\bibnamefont {Guillam{\'{o}}n}}, \bibinfo {author} {\bibfnamefont {H.}~\bibnamefont {Suderow}}, \bibinfo {author} {\bibfnamefont {J.~I.}\ \bibnamefont {Pascual}}, \ and\ \bibinfo {author} {\bibfnamefont {N.}~\bibnamefont {Lorente}},\ }\href {\doibase 10.1103/PhysRevLett.120.167001} {\bibfield  {journal} {\bibinfo  {journal} {Phys. Rev. Lett.}\ }\textbf {\bibinfo {volume} {120}},\ \bibinfo {pages} {1} (\bibinfo {year} {2018})}\BibitemShut {NoStop}%
\bibitem [{\citenamefont {Trivini}\ \emph {et~al.}(2024)\citenamefont {Trivini}, \citenamefont {Ortuzar}, \citenamefont {Zaldivar}, \citenamefont {Herrera}, \citenamefont {Guillam{\'o}n}, \citenamefont {Suderow}, \citenamefont {Bergeret},\ and\ \citenamefont {Pascual}}]{Trivini2024}%
  \BibitemOpen
  \bibfield  {author} {\bibinfo {author} {\bibfnamefont {S.}~\bibnamefont {Trivini}}, \bibinfo {author} {\bibfnamefont {J.}~\bibnamefont {Ortuzar}}, \bibinfo {author} {\bibfnamefont {J.}~\bibnamefont {Zaldivar}}, \bibinfo {author} {\bibfnamefont {E.}~\bibnamefont {Herrera}}, \bibinfo {author} {\bibfnamefont {I.}~\bibnamefont {Guillam{\'o}n}}, \bibinfo {author} {\bibfnamefont {H.}~\bibnamefont {Suderow}}, \bibinfo {author} {\bibfnamefont {F.~S.}\ \bibnamefont {Bergeret}}, \ and\ \bibinfo {author} {\bibfnamefont {J.~I.}\ \bibnamefont {Pascual}},\ }\href {\doibase 10.1103/PhysRevB.110.235405} {\bibfield  {journal} {\bibinfo  {journal} {Physical Review B}\ }\textbf {\bibinfo {volume} {110}},\ \bibinfo {pages} {235405} (\bibinfo {year} {2024})}\BibitemShut {NoStop}%
\bibitem [{\citenamefont {Pan}\ \emph {et~al.}(1998)\citenamefont {Pan}, \citenamefont {Hudson},\ and\ \citenamefont {Davis}}]{Pan1998}%
  \BibitemOpen
  \bibfield  {author} {\bibinfo {author} {\bibfnamefont {S.~H.}\ \bibnamefont {Pan}}, \bibinfo {author} {\bibfnamefont {E.~W.}\ \bibnamefont {Hudson}}, \ and\ \bibinfo {author} {\bibfnamefont {J.~C.}\ \bibnamefont {Davis}},\ }\href {\doibase 10.1063/1.122654} {\bibfield  {journal} {\bibinfo  {journal} {Appl. Phys. Lett.}\ }\textbf {\bibinfo {volume} {73}},\ \bibinfo {pages} {2992} (\bibinfo {year} {1998})}\BibitemShut {NoStop}%
\bibitem [{\citenamefont {Suderow}\ \emph {et~al.}(2002)\citenamefont {Suderow}, \citenamefont {Crespo}, \citenamefont {Martinez-Samper}, \citenamefont {Rodrigo}, \citenamefont {Rubio-Bollinger}, \citenamefont {Vieira}, \citenamefont {Luchier}, \citenamefont {Brison},\ and\ \citenamefont {Canfield}}]{Suderow2002}%
  \BibitemOpen
  \bibfield  {author} {\bibinfo {author} {\bibfnamefont {H.}~\bibnamefont {Suderow}}, \bibinfo {author} {\bibfnamefont {M.}~\bibnamefont {Crespo}}, \bibinfo {author} {\bibfnamefont {P.}~\bibnamefont {Martinez-Samper}}, \bibinfo {author} {\bibfnamefont {J.~G.}\ \bibnamefont {Rodrigo}}, \bibinfo {author} {\bibfnamefont {G.}~\bibnamefont {Rubio-Bollinger}}, \bibinfo {author} {\bibfnamefont {S.}~\bibnamefont {Vieira}}, \bibinfo {author} {\bibfnamefont {N.}~\bibnamefont {Luchier}}, \bibinfo {author} {\bibfnamefont {J.~P.}\ \bibnamefont {Brison}}, \ and\ \bibinfo {author} {\bibfnamefont {P.~C.}\ \bibnamefont {Canfield}},\ }in\ \href {\doibase 10.1016/S0921-4534(01)01228-X} {\emph {\bibinfo {booktitle} {Phys. C Supercond. its Appl.}}},\ Vol.\ \bibinfo {volume} {369}\ (\bibinfo  {publisher} {North-Holland},\ \bibinfo {year} {2002})\ pp.\ \bibinfo {pages} {106--112}\BibitemShut {NoStop}%
\bibitem [{\citenamefont {Rodrigo}\ \emph {et~al.}(2004)\citenamefont {Rodrigo}, \citenamefont {Suderow},\ and\ \citenamefont {Vieira}}]{Rodrigo2004}%
  \BibitemOpen
  \bibfield  {author} {\bibinfo {author} {\bibfnamefont {J.~G.}\ \bibnamefont {Rodrigo}}, \bibinfo {author} {\bibfnamefont {H.}~\bibnamefont {Suderow}}, \ and\ \bibinfo {author} {\bibfnamefont {S.}~\bibnamefont {Vieira}},\ }\href {\doibase 10.1140/EPJB/E2004-00273-Y} {\bibfield  {journal} {\bibinfo  {journal} {Eur. Phys. J. B}\ }\textbf {\bibinfo {volume} {40}},\ \bibinfo {pages} {483} (\bibinfo {year} {2004})}\BibitemShut {NoStop}%
\bibitem [{\citenamefont {Franke}\ \emph {et~al.}(2011)\citenamefont {Franke}, \citenamefont {Schulze},\ and\ \citenamefont {Pascual}}]{Franke2011}%
  \BibitemOpen
  \bibfield  {author} {\bibinfo {author} {\bibfnamefont {K.~J.}\ \bibnamefont {Franke}}, \bibinfo {author} {\bibfnamefont {G.}~\bibnamefont {Schulze}}, \ and\ \bibinfo {author} {\bibfnamefont {J.~I.}\ \bibnamefont {Pascual}},\ }\href {\doibase 10.1126/science.1202204} {\bibfield  {journal} {\bibinfo  {journal} {Science}\ }\textbf {\bibinfo {volume} {332}},\ \bibinfo {pages} {940} (\bibinfo {year} {2011})}\BibitemShut {NoStop}%
\bibitem [{\citenamefont {Ruby}\ \emph {et~al.}(2016)\citenamefont {Ruby}, \citenamefont {Peng}, \citenamefont {von Oppen}, \citenamefont {Heinrich},\ and\ \citenamefont {Franke}}]{Ruby2016}%
  \BibitemOpen
  \bibfield  {author} {\bibinfo {author} {\bibfnamefont {M.}~\bibnamefont {Ruby}}, \bibinfo {author} {\bibfnamefont {Y.}~\bibnamefont {Peng}}, \bibinfo {author} {\bibfnamefont {F.}~\bibnamefont {von Oppen}}, \bibinfo {author} {\bibfnamefont {B.~W.}\ \bibnamefont {Heinrich}}, \ and\ \bibinfo {author} {\bibfnamefont {K.~J.}\ \bibnamefont {Franke}},\ }\href {\doibase 10.1103/PhysRevLett.117.186801} {\bibfield  {journal} {\bibinfo  {journal} {Phys. Rev. Lett.}\ }\textbf {\bibinfo {volume} {117}},\ \bibinfo {pages} {186801} (\bibinfo {year} {2016})}\BibitemShut {NoStop}%
\bibitem [{\citenamefont {Choi}\ \emph {et~al.}(2017)\citenamefont {Choi}, \citenamefont {Rubio-Verd{\'{u}}}, \citenamefont {de~Bruijckere}, \citenamefont {Ugeda}, \citenamefont {Lorente},\ and\ \citenamefont {Pascual}}]{Choi2017}%
  \BibitemOpen
  \bibfield  {author} {\bibinfo {author} {\bibfnamefont {D.-J.}\ \bibnamefont {Choi}}, \bibinfo {author} {\bibfnamefont {C.}~\bibnamefont {Rubio-Verd{\'{u}}}}, \bibinfo {author} {\bibfnamefont {J.}~\bibnamefont {de~Bruijckere}}, \bibinfo {author} {\bibfnamefont {M.~M.}\ \bibnamefont {Ugeda}}, \bibinfo {author} {\bibfnamefont {N.}~\bibnamefont {Lorente}}, \ and\ \bibinfo {author} {\bibfnamefont {J.~I.}\ \bibnamefont {Pascual}},\ }\href {\doibase 10.1038/ncomms15175} {\bibfield  {journal} {\bibinfo  {journal} {Nat. Commun.}\ }\textbf {\bibinfo {volume} {8}},\ \bibinfo {pages} {15175} (\bibinfo {year} {2017})}\BibitemShut {NoStop}%
\bibitem [{\citenamefont {Ka{\v{c}}mar{\v{c}}{\'{i}}k}\ \emph {et~al.}(2016)\citenamefont {Ka{\v{c}}mar{\v{c}}{\'{i}}k}, \citenamefont {Pribulov{\'{a}}}, \citenamefont {Samuely}, \citenamefont {Szab{\'{o}}}, \citenamefont {Cambel}, \citenamefont {{\v{S}}olt{\'{y}}s}, \citenamefont {Herrera}, \citenamefont {Suderow}, \citenamefont {Correa-Orellana}, \citenamefont {Prabhakaran},\ and\ \citenamefont {Samuely}}]{Kacmarcik2016a}%
  \BibitemOpen
  \bibfield  {author} {\bibinfo {author} {\bibfnamefont {J.}~\bibnamefont {Ka{\v{c}}mar{\v{c}}{\'{i}}k}}, \bibinfo {author} {\bibfnamefont {Z.}~\bibnamefont {Pribulov{\'{a}}}}, \bibinfo {author} {\bibfnamefont {T.}~\bibnamefont {Samuely}}, \bibinfo {author} {\bibfnamefont {P.}~\bibnamefont {Szab{\'{o}}}}, \bibinfo {author} {\bibfnamefont {V.}~\bibnamefont {Cambel}}, \bibinfo {author} {\bibfnamefont {J.}~\bibnamefont {{\v{S}}olt{\'{y}}s}}, \bibinfo {author} {\bibfnamefont {E.}~\bibnamefont {Herrera}}, \bibinfo {author} {\bibfnamefont {H.}~\bibnamefont {Suderow}}, \bibinfo {author} {\bibfnamefont {A.}~\bibnamefont {Correa-Orellana}}, \bibinfo {author} {\bibfnamefont {D.}~\bibnamefont {Prabhakaran}}, \ and\ \bibinfo {author} {\bibfnamefont {P.}~\bibnamefont {Samuely}},\ }\href {\doibase 10.1103/PhysRevB.93.144502} {\bibfield  {journal} {\bibinfo  {journal} {Physical Review B}\ }\textbf {\bibinfo {volume} {93}},\ \bibinfo {pages} {144502} (\bibinfo {year} {2016})}\BibitemShut {NoStop}%
\bibitem [{\citenamefont {Prist\'a\ifmmode~\check{s}\else \v{s}\fi{}}\ \emph {et~al.}(2018)\citenamefont {Prist\'a\ifmmode~\check{s}\else \v{s}\fi{}}, \citenamefont {Orend\'a\ifmmode~\check{c}\else \v{c}\fi{}}, \citenamefont {Gab\'ani}, \citenamefont {Ka\ifmmode \check{c}\else \v{c}\fi{}mar\ifmmode~\check{c}\else \v{c}\fi{}\'{\i}k}, \citenamefont {Ga\ifmmode~\check{z}\else \v{z}\fi{}o}, \citenamefont {Pribulov\'a}, \citenamefont {Correa-Orellana}, \citenamefont {Herrera}, \citenamefont {Suderow},\ and\ \citenamefont {Samuely}}]{pristas2018}%
  \BibitemOpen
  \bibfield  {author} {\bibinfo {author} {\bibfnamefont {G.}~\bibnamefont {Prist\'a\ifmmode~\check{s}\else \v{s}\fi{}}}, \bibinfo {author} {\bibfnamefont {M.}~\bibnamefont {Orend\'a\ifmmode~\check{c}\else \v{c}\fi{}}}, \bibinfo {author} {\bibfnamefont {S.}~\bibnamefont {Gab\'ani}}, \bibinfo {author} {\bibfnamefont {J.}~\bibnamefont {Ka\ifmmode \check{c}\else \v{c}\fi{}mar\ifmmode~\check{c}\else \v{c}\fi{}\'{\i}k}}, \bibinfo {author} {\bibfnamefont {E.}~\bibnamefont {Ga\ifmmode~\check{z}\else \v{z}\fi{}o}}, \bibinfo {author} {\bibfnamefont {Z.}~\bibnamefont {Pribulov\'a}}, \bibinfo {author} {\bibfnamefont {A.}~\bibnamefont {Correa-Orellana}}, \bibinfo {author} {\bibfnamefont {E.}~\bibnamefont {Herrera}}, \bibinfo {author} {\bibfnamefont {H.}~\bibnamefont {Suderow}}, \ and\ \bibinfo {author} {\bibfnamefont {P.}~\bibnamefont {Samuely}},\ }\href {\doibase 10.1103/PhysRevB.97.134505} {\bibfield  {journal} {\bibinfo  {journal} {Phys. Rev. B}\ }\textbf {\bibinfo {volume} {97}},\ \bibinfo {pages} {134505} (\bibinfo
  {year} {2018})}\BibitemShut {NoStop}%
\bibitem [{\citenamefont {Che}\ \emph {et~al.}(2016)\citenamefont {Che}, \citenamefont {Le}, \citenamefont {Xu}, \citenamefont {Xing}, \citenamefont {Shi}, \citenamefont {Xu},\ and\ \citenamefont {Lu}}]{Che2016}%
  \BibitemOpen
  \bibfield  {author} {\bibinfo {author} {\bibfnamefont {L.}~\bibnamefont {Che}}, \bibinfo {author} {\bibfnamefont {T.}~\bibnamefont {Le}}, \bibinfo {author} {\bibfnamefont {C.~Q.}\ \bibnamefont {Xu}}, \bibinfo {author} {\bibfnamefont {X.~Z.}\ \bibnamefont {Xing}}, \bibinfo {author} {\bibfnamefont {Z.}~\bibnamefont {Shi}}, \bibinfo {author} {\bibfnamefont {X.}~\bibnamefont {Xu}}, \ and\ \bibinfo {author} {\bibfnamefont {X.}~\bibnamefont {Lu}},\ }\href {\doibase 10.1103/PhysRevB.94.024519} {\bibfield  {journal} {\bibinfo  {journal} {Phys. Rev. B}\ }\textbf {\bibinfo {volume} {94}},\ \bibinfo {pages} {024519} (\bibinfo {year} {2016})}\BibitemShut {NoStop}%
\bibitem [{\citenamefont {M{\'e}nard}\ \emph {et~al.}(2015)\citenamefont {M{\'e}nard}, \citenamefont {Guissart}, \citenamefont {Brun}, \citenamefont {Pons}, \citenamefont {Stolyarov}, \citenamefont {Debontridder}, \citenamefont {Leclerc}, \citenamefont {Janod}, \citenamefont {Cario}, \citenamefont {Roditchev}, \citenamefont {Simon},\ and\ \citenamefont {Cren}}]{Menard2015}%
  \BibitemOpen
  \bibfield  {author} {\bibinfo {author} {\bibfnamefont {G.~C.}\ \bibnamefont {M{\'e}nard}}, \bibinfo {author} {\bibfnamefont {S.}~\bibnamefont {Guissart}}, \bibinfo {author} {\bibfnamefont {C.}~\bibnamefont {Brun}}, \bibinfo {author} {\bibfnamefont {S.}~\bibnamefont {Pons}}, \bibinfo {author} {\bibfnamefont {V.~S.}\ \bibnamefont {Stolyarov}}, \bibinfo {author} {\bibfnamefont {F.}~\bibnamefont {Debontridder}}, \bibinfo {author} {\bibfnamefont {M.~V.}\ \bibnamefont {Leclerc}}, \bibinfo {author} {\bibfnamefont {E.}~\bibnamefont {Janod}}, \bibinfo {author} {\bibfnamefont {L.}~\bibnamefont {Cario}}, \bibinfo {author} {\bibfnamefont {D.}~\bibnamefont {Roditchev}}, \bibinfo {author} {\bibfnamefont {P.}~\bibnamefont {Simon}}, \ and\ \bibinfo {author} {\bibfnamefont {T.}~\bibnamefont {Cren}},\ }\href {\doibase 10.1038/nphys3508} {\bibfield  {journal} {\bibinfo  {journal} {Nature Physics}\ }\textbf {\bibinfo {volume} {11}},\ \bibinfo {pages} {1013} (\bibinfo {year} {2015})}\BibitemShut {NoStop}%
\bibitem [{\citenamefont {Kim}\ \emph {et~al.}(2020)\citenamefont {Kim}, \citenamefont {R{\'{o}}zsa}, \citenamefont {Schreyer}, \citenamefont {Simon},\ and\ \citenamefont {Wiesendanger}}]{Kim2020}%
  \BibitemOpen
  \bibfield  {author} {\bibinfo {author} {\bibfnamefont {H.}~\bibnamefont {Kim}}, \bibinfo {author} {\bibfnamefont {L.}~\bibnamefont {R{\'{o}}zsa}}, \bibinfo {author} {\bibfnamefont {D.}~\bibnamefont {Schreyer}}, \bibinfo {author} {\bibfnamefont {E.}~\bibnamefont {Simon}}, \ and\ \bibinfo {author} {\bibfnamefont {R.}~\bibnamefont {Wiesendanger}},\ }\href {\doibase 10.1038/s41467-020-18406-8} {\bibfield  {journal} {\bibinfo  {journal} {Nat. Commun.}\ }\textbf {\bibinfo {volume} {11}},\ \bibinfo {pages} {1} (\bibinfo {year} {2020})}\BibitemShut {NoStop}%
\bibitem [{\citenamefont {Ortuzar}\ \emph {et~al.}(2022)\citenamefont {Ortuzar}, \citenamefont {Trivini}, \citenamefont {Alvarado}, \citenamefont {Rouco}, \citenamefont {Zaldivar}, \citenamefont {Yeyati}, \citenamefont {Pascual},\ and\ \citenamefont {Bergeret}}]{ortuzar2022}%
  \BibitemOpen
  \bibfield  {author} {\bibinfo {author} {\bibfnamefont {J.}~\bibnamefont {Ortuzar}}, \bibinfo {author} {\bibfnamefont {S.}~\bibnamefont {Trivini}}, \bibinfo {author} {\bibfnamefont {M.}~\bibnamefont {Alvarado}}, \bibinfo {author} {\bibfnamefont {M.}~\bibnamefont {Rouco}}, \bibinfo {author} {\bibfnamefont {J.}~\bibnamefont {Zaldivar}}, \bibinfo {author} {\bibfnamefont {A.~L.}\ \bibnamefont {Yeyati}}, \bibinfo {author} {\bibfnamefont {J.~I.}\ \bibnamefont {Pascual}}, \ and\ \bibinfo {author} {\bibfnamefont {F.~S.}\ \bibnamefont {Bergeret}},\ }\href {\doibase 10.1103/PhysRevB.105.245403} {\bibfield  {journal} {\bibinfo  {journal} {Phys. Rev. B}\ }\textbf {\bibinfo {volume} {105}},\ \bibinfo {pages} {245403} (\bibinfo {year} {2022})}\BibitemShut {NoStop}%
\bibitem [{\citenamefont {Uldemolins}\ \emph {et~al.}(2022)\citenamefont {Uldemolins}, \citenamefont {Mesaros},\ and\ \citenamefont {Simon}}]{uldemolins2022}%
  \BibitemOpen
  \bibfield  {author} {\bibinfo {author} {\bibfnamefont {M.}~\bibnamefont {Uldemolins}}, \bibinfo {author} {\bibfnamefont {A.}~\bibnamefont {Mesaros}}, \ and\ \bibinfo {author} {\bibfnamefont {P.}~\bibnamefont {Simon}},\ }\href {\doibase 10.1103/PhysRevB.105.144503} {\bibfield  {journal} {\bibinfo  {journal} {Phys. Rev. B}\ }\textbf {\bibinfo {volume} {105}},\ \bibinfo {pages} {144503} (\bibinfo {year} {2022})}\BibitemShut {NoStop}%
\bibitem [{\citenamefont {Bergeret}\ \emph {et~al.}(2005)\citenamefont {Bergeret}, \citenamefont {Volkov},\ and\ \citenamefont {Efetov}}]{Bergeret2005}%
  \BibitemOpen
  \bibfield  {author} {\bibinfo {author} {\bibfnamefont {F.~S.}\ \bibnamefont {Bergeret}}, \bibinfo {author} {\bibfnamefont {A.~F.}\ \bibnamefont {Volkov}}, \ and\ \bibinfo {author} {\bibfnamefont {K.~B.}\ \bibnamefont {Efetov}},\ }\href {\doibase 10.1103/RevModPhys.77.1321} {\bibfield  {journal} {\bibinfo  {journal} {Rev. Mod. Phys.}\ }\textbf {\bibinfo {volume} {77}},\ \bibinfo {pages} {1321} (\bibinfo {year} {2005})}\BibitemShut {NoStop}%
\bibitem [{\citenamefont {Bergeret}\ and\ \citenamefont {Volkov}(2023)}]{bergeret2023triplet}%
  \BibitemOpen
  \bibfield  {author} {\bibinfo {author} {\bibfnamefont {F.~S.}\ \bibnamefont {Bergeret}}\ and\ \bibinfo {author} {\bibfnamefont {A.~F.}\ \bibnamefont {Volkov}},\ }\href@noop {} {\bibfield  {journal} {\bibinfo  {journal} {Annals of Physics}\ }\textbf {\bibinfo {volume} {456}},\ \bibinfo {pages} {169232} (\bibinfo {year} {2023})}\BibitemShut {NoStop}%
\bibitem [{\citenamefont {Kanasugi}\ and\ \citenamefont {Yanase}(2022)}]{kanasugi2022}%
  \BibitemOpen
  \bibfield  {author} {\bibinfo {author} {\bibfnamefont {S.}~\bibnamefont {Kanasugi}}\ and\ \bibinfo {author} {\bibfnamefont {Y.}~\bibnamefont {Yanase}},\ }\href {\doibase 10.1038/s42005-022-00804-7} {\bibfield  {journal} {\bibinfo  {journal} {Commun Phys}\ }\textbf {\bibinfo {volume} {5}},\ \bibinfo {pages} {39} (\bibinfo {year} {2022})}\BibitemShut {NoStop}%
\bibitem [{\citenamefont {Fulde}\ and\ \citenamefont {Ferrell}(1964)}]{Fulde1964}%
  \BibitemOpen
  \bibfield  {author} {\bibinfo {author} {\bibfnamefont {P.}~\bibnamefont {Fulde}}\ and\ \bibinfo {author} {\bibfnamefont {R.~A.}\ \bibnamefont {Ferrell}},\ }\href {\doibase 10.1103/PhysRev.135.A550} {\bibfield  {journal} {\bibinfo  {journal} {Physical Review}\ }\textbf {\bibinfo {volume} {135}},\ \bibinfo {pages} {A550–A563} (\bibinfo {year} {1964})}\BibitemShut {NoStop}%
\bibitem [{\citenamefont {Larkin}(1965)}]{larkin1965inhomogeneous}%
  \BibitemOpen
  \bibfield  {author} {\bibinfo {author} {\bibfnamefont {A.~I.}\ \bibnamefont {Larkin}},\ }\href@noop {} {\bibfield  {journal} {\bibinfo  {journal} {Sov. Phys. JETP}\ }\textbf {\bibinfo {volume} {20}},\ \bibinfo {pages} {762} (\bibinfo {year} {1965})}\BibitemShut {NoStop}%
\bibitem [{\citenamefont {Yuan}\ and\ \citenamefont {Fu}(2022)}]{Yuan2022}%
  \BibitemOpen
  \bibfield  {author} {\bibinfo {author} {\bibfnamefont {N.~F.~Q.}\ \bibnamefont {Yuan}}\ and\ \bibinfo {author} {\bibfnamefont {L.}~\bibnamefont {Fu}},\ }\href {\doibase 10.1073/pnas.2119548119} {\bibfield  {journal} {\bibinfo  {journal} {Proceedings of the National Academy of Sciences}\ }\textbf {\bibinfo {volume} {119}},\ \bibinfo {pages} {e2119548119} (\bibinfo {year} {2022})}\BibitemShut {NoStop}%
\bibitem [{\citenamefont {Ili{\'c}}\ and\ \citenamefont {Bergeret}(2022)}]{ilic2022theory}%
  \BibitemOpen
  \bibfield  {author} {\bibinfo {author} {\bibfnamefont {S.}~\bibnamefont {Ili{\'c}}}\ and\ \bibinfo {author} {\bibfnamefont {F.~S.}\ \bibnamefont {Bergeret}},\ }\href@noop {} {\bibfield  {journal} {\bibinfo  {journal} {Physical Review Letters}\ }\textbf {\bibinfo {volume} {128}},\ \bibinfo {pages} {177001} (\bibinfo {year} {2022})}\BibitemShut {NoStop}%
\bibitem [{\citenamefont {Powell}\ \emph {et~al.}(2025)\citenamefont {Powell}, \citenamefont {Kuang}, \citenamefont {{Hawkins-Pottier}}, \citenamefont {Jalil}, \citenamefont {Birkbeck}, \citenamefont {Jiang}, \citenamefont {Kim}, \citenamefont {Zou}, \citenamefont {Komrakova}, \citenamefont {Haigh}, \citenamefont {Timokhin}, \citenamefont {Balakrishnan}, \citenamefont {Geim}, \citenamefont {Walet}, \citenamefont {Principi},\ and\ \citenamefont {Grigorieva}}]{Powell2025}%
  \BibitemOpen
  \bibfield  {author} {\bibinfo {author} {\bibfnamefont {L.}~\bibnamefont {Powell}}, \bibinfo {author} {\bibfnamefont {W.}~\bibnamefont {Kuang}}, \bibinfo {author} {\bibfnamefont {G.}~\bibnamefont {{Hawkins-Pottier}}}, \bibinfo {author} {\bibfnamefont {R.}~\bibnamefont {Jalil}}, \bibinfo {author} {\bibfnamefont {J.}~\bibnamefont {Birkbeck}}, \bibinfo {author} {\bibfnamefont {Z.}~\bibnamefont {Jiang}}, \bibinfo {author} {\bibfnamefont {M.}~\bibnamefont {Kim}}, \bibinfo {author} {\bibfnamefont {Y.}~\bibnamefont {Zou}}, \bibinfo {author} {\bibfnamefont {S.}~\bibnamefont {Komrakova}}, \bibinfo {author} {\bibfnamefont {S.}~\bibnamefont {Haigh}}, \bibinfo {author} {\bibfnamefont {I.}~\bibnamefont {Timokhin}}, \bibinfo {author} {\bibfnamefont {G.}~\bibnamefont {Balakrishnan}}, \bibinfo {author} {\bibfnamefont {A.~K.}\ \bibnamefont {Geim}}, \bibinfo {author} {\bibfnamefont {N.}~\bibnamefont {Walet}}, \bibinfo {author} {\bibfnamefont {A.}~\bibnamefont {Principi}}, \ and\ \bibinfo {author} {\bibfnamefont {I.~V.}\
  \bibnamefont {Grigorieva}},\ }\href {\doibase 10.1038/s41467-024-54867-x} {\bibfield  {journal} {\bibinfo  {journal} {Nature Communications}\ }\textbf {\bibinfo {volume} {16}},\ \bibinfo {pages} {291} (\bibinfo {year} {2025})}\BibitemShut {NoStop}%
\end{thebibliography}

%

\end{document}


\title{Revealing inter-band  electron pairing in a superconductor with spin-orbit coupling}

\author{Javier Zald\'ivar}  \thanks{These two authors contributed equally}
  \affiliation{CIC nanoGUNE-BRTA, 20018 Donostia-San Sebasti\'an, Spain}

\author{Jon Ortuzar}  \thanks{These two authors contributed equally}
  \affiliation{CIC nanoGUNE-BRTA, 20018 Donostia-San Sebasti\'an, Spain}
  
  \author{Miguel Alvarado}
  \affiliation{ 
Instituto de Ciencia de Materiales de Madrid (ICMM), Sor Juana In{\'e}s de la Cruz 3, 28049 Madrid, Spain}
  \affiliation{Departamento de Física Teórica de la Materia Condensada and Condensed Matter Physics Center (IFIMAC), Universidad Autónoma de Madrid, Madrid, Spain }

\author{Stefano Trivini}
  \affiliation{CIC nanoGUNE-BRTA, 20018 Donostia-San Sebasti\'an, Spain}

\author{Julie Baumard}
    \affiliation{Centro de Física de Materiales (CFM-MPC), Centro Mixto CSIC-UPV/EHU, 20018 San Sebastián, Spain}
  
\author{Carmen Rubio-Verd\'u }
  \affiliation{ICFO, 08860 Castelldefels, Barcelona, Spain}

\author{Edwin Herrera}
  \affiliation{Departamento de Física Teórica de la Materia Condensada and Condensed Matter Physics Center (IFIMAC), Universidad Autónoma de Madrid, Madrid, Spain }
  
\author{Hermann Suderow}
  \affiliation{Departamento de Física Teórica de la Materia Condensada and Condensed Matter Physics Center (IFIMAC), Universidad Autónoma de Madrid, Madrid, Spain }

\author{Alfredo Levy Yeyati}
  \affiliation{Departamento de Física Teórica de la Materia Condensada and Condensed Matter Physics Center (IFIMAC), Universidad Autónoma de Madrid, Madrid, Spain }

\author{F. Sebastian Bergeret}
    \affiliation{Centro de Física de Materiales (CFM-MPC), Centro Mixto CSIC-UPV/EHU, 20018 San Sebastián, Spain}
	\affiliation{Donostia International Physics Center (DIPC), 20018 Donostia-San Sebasti\'an, Spain}
  
\author{Jose Ignacio Pascual}
  \affiliation{CIC nanoGUNE-BRTA, 20018 Donostia-San Sebasti\'an, Spain}
\affiliation{Ikerbasque, Basque Foundation for Science, Bilbao, Spain}

\renewcommand{\abstractname}{\vspace{1cm} }	
\date{\today}
\begin{abstract}
\baselineskip12pt
	\tableofcontents		
\end{abstract}	
\baselineskip14pt
\maketitle 

\newpage

\section{Experimental methods}\label{SN:methods}

The experiments were conducted on the clean Bi-terminated surface of $\beta$-Bi$_2$Pd(001) samples, grown following the procedure described in \cite{Herrera2015}. The crystals were cleaved at approximately 100 K under ultra-high vacuum (UHV) conditions (P\textless 10$^{-10}$ mbar) just before being transferred to a low-temperature scanning tunneling microscope (JT-SPECS GmbH) with a base temperature of 1.3 K. The $\beta$-Bi$2$Pd substrate becomes superconducting below a critical temperature of 4.5 K, exhibiting a superconducting gap of $\Delta_{sample}$ = 0.775 meV at this temperature. 

All measurements were performed using a \bipd\ superconducting tip. A tungsten (W) tip was in situ coated with superconducting substrate material by repeated indentations onto the clean $\beta$-Bi$2$Pd(001) surface until the width of the superconducting gap in the tunneling spectra doubled. This confirmed the formation of a symmetric Superconductor-Insulator-Superconductor (SIS) junction between the tip and the sample, enabling an energy resolution below 0.06 meV in the spectra. In an SIS junction, the spectroscopic features of the substrate are shifted by the superconducting gap of the tip, which in our results is $\Delta_{tip}$ $\sim$ 0.775 meV (see below).

Vanadium adatoms, sourced from a 99.8 $\%$ purity rod (Goodfellow Inc.), were deposited directly onto the sample in the STM stage at temperatures below 15 K. Several coverage levels were explored throughout the experiment to rule out-coupling effects between magnetic impurities \cite{Trivini2024}.

\section{Quasiparticle Interference of \bipd} \label{SN-QPI}

To study quasiparticle interference (QPI) in the main manuscript, we perform Fast Fourier Transformation (FFT) of dI/dV images of the \bipd\ sample obtained at a fixed bias voltage value applied to the substrate of the STM. Using bias values to tunnel well above the superconducting gap allows us to explore the QPI of electronic bands in the normal metal density of states with all kinds of scattering elements at the surface. Using sub-gap voltage bias, we explore the scattering of  Bogoliubov quasiparticles bound to magnetic impurities exclusively, thus involving only bands hybridized with the magnetic impurity. 

\begin{figure*}[ht]
  \includegraphics[width=0.9\textwidth]{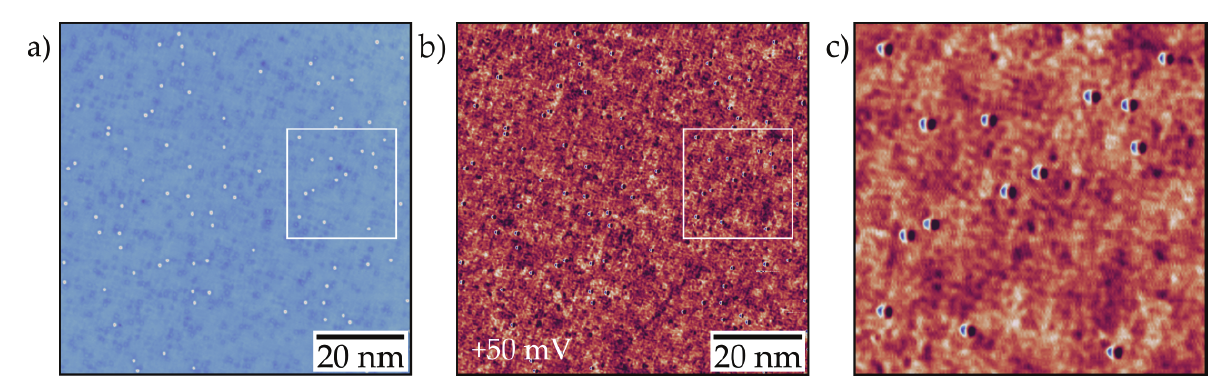}
  \caption{Quasiparticle Interference at +50 mV. a) Topographic map of an 80x80 nm
area showing the amount and distribution of defects. b) Simultaneously-acquired dI/dV
map showing the modulations associated with QPI. (V=50 mV, I=500 pA, $V_{RMS}$=100 $\mu$V).
c) Zoom into the area marked by a white square, showing the modulations of dI/dV
around defects. The modulations are homogeneously distributed along the area.} \label{SF-QPI0}
\end{figure*}

\textbf{Normal state QPI}: The quasiparticle interference (QPI) patterns outside the superconducting gap of \bipd\ were previously investigated in Ref.\cite{Iwaya2017}. In our study, we performed large-area constant-current dI/dV mapping on the bare substrate, as shown in Fig.\ref{SF-QPI0}a. The presence of surface and subsurface vacancies, along with bismuth adatoms, induces local density of states (LDoS) modulations across the surface, which are captured in maps such as those in Figs.~\ref{SF-QPI0}b and \ref{SF-QPI0}c.
Since no magnetic impurities are present, the observed QPI arises from finite potential scattering caused by defects and bismuth adatoms. 

\begin{figure*}[ht]
  \includegraphics[width=0.9\textwidth]{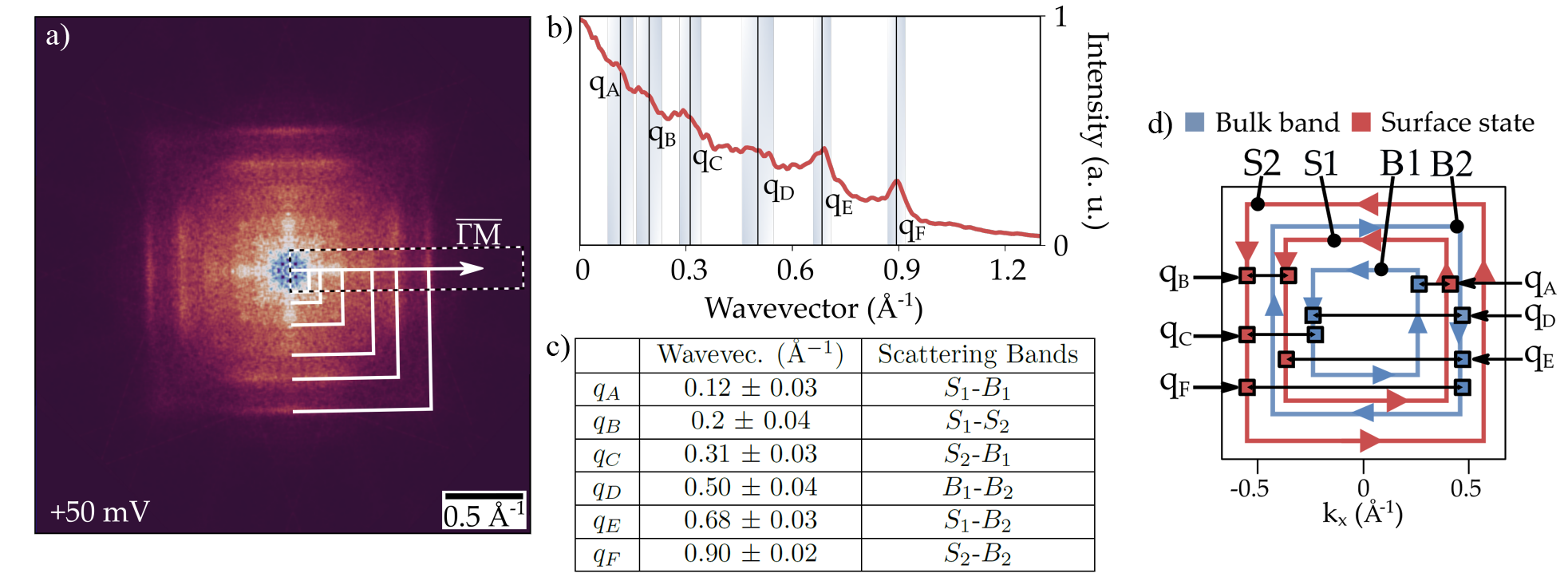}
  \caption{Estimation of scattering wavevectors from FFT maps a) 4-fold symmetrized 2D-FastFourier Transform of a dI/dV map measured at +50 mV. Each square corresponds to a different scattering vector. b) Accumulated intensity in the area marked by a dotted square around the $\overline{\Gamma \mbox{X}}$ axis in a). Vertical lines indicate the scattering vectors and shaded areas indicate the estimation error. c) Table of scattering wavevectors obtained from b), and identification of interband transitions, follwong. } \label{SF-QPI}
\end{figure*}

The Fast Fourier Transform (FFT) of these dI/dV maps reveals four-fold symmetric contours, consistent with the band structure of \bipd. To enhance weak signals, we applied a four-fold symmetrization procedure by averaging the maps over four 90$^{\circ}$ rotations. This approach results in FFT maps such as the one shown in Fig.~\ref{SF-QPI}a.
Additionally, the resulting scattering vectors can be extracted from line cuts of the FFT intensity, as illustrated in Fig.~\ref{SF-QPI}b.

Six possible concentric contours can be resolved, attributed to the six spin-conserving scattering vectors connecting the four spin non-degenerate electronic bands of the bare \bipd\ substrate, two surface states and two projected bulk bands, as described in the main manuscript.  The values of the scattering vectors are summarized in Fig. \ref{SF-QPI} (c). As discussed in the manuscript, the magnitude of these scattering vectors agrees well with inter-band scattering vectors following results from the band structure of \bipd\ obtained from spin-polarized ARPES measurements and DFT simulations from Refs.~\cite{Iwaya2017,Sakano2015}.  \\
\\

\newpage

\textbf{Scattering of YSR Bogoliubov quasiparticles}: Oscillations of YSR density of states around magnetic impurities have been observed previously in Refs.~\cite{Ruby2016,menard2015,kim2020} and others, reflecting QPI of Bogoliubov quasiparticles (BQPI). Due to the subgap nature of these excitations, dI/dV maps had to be measured with the feedback loop disconnected. To avoid dragging V and Bi adatoms, we used a dual pass technique:  we recorded first a topographic scan line with bias outside the gap and the feedback loop closed, and then repeated the scan line with the feedback off, using the recorded height profile at the subgap bias voltage of a YSR state. 

 \begin{figure*}[tb]
  \includegraphics[width=0.8\textwidth]{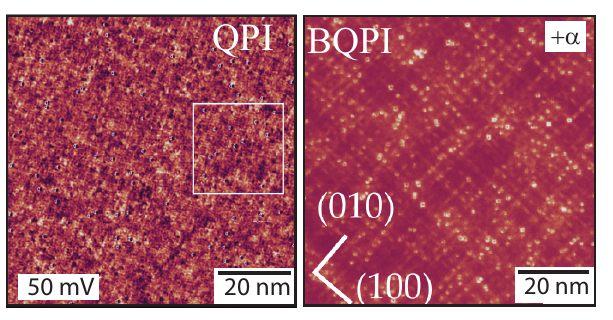}
  \caption{Differential conductance maps measured in constant current (QPI) and dual pass (BQPI) modes at 50 mV and 1 mV (i.e. below the onset of the superconducting gap edge at 1.5 meV), respectively. 
  } \label{SF:QPI2}
\end{figure*}

In Figure \ref{SF:QPI2}, we compare the BQPI pattern of a \bipd\ sample with V adatoms, measured at the sub-gap  bias of the $\alpha_+$ YSR state, with the normal state QPI pattern of Fig.~\ref{SF-QPI} (i.e. in the absence of magnetic impurities, although the presence of V impurities does not bring any change). BQPI patterns appears only around V impurities and along the main directions of the surface, in contrast to the homogeneous distribution of the LDOS oscillations outside the superconducting. As discussed in the manuscript, in BQPI only electronic bands at the \bipd\ surface that hybridized with the V impurity can participate in subgap quasiparticle scattering.  
 
\newpage
\section{Orbital character of YSR states on \bipd} \label{SN-orbital}

YSR states originate from the exchange interaction between the superconductor and spin-polarized localized orbitals of a magnetic impurity. The spin of free vanadium atoms is S=3/2 atom due to three singly occupied d-orbitals. 
On the surface, the interaction of each of these singly occupied $d$ orbital (with spin $\bm{S_n}$) with electron spins $\bm{\sigma_k}$ of the $ith$ substrate's band $\bm{k_i}$ is described by a magnetic exchange term $J_{n,i}\bm{S_n\cdot \sigma_{k_i}}$, where $J_{n,i}$, the exchange amplitude, determines the energy $\epsilon_n$ ($n=\alpha, \beta, \gamma$) of the YSR state.  Variations of wavefunction overlap of $d$ and $\bm{k_i}$ may result in the different exchange coupling strengths $J_{n,i}$ for each channel and, consequently, YSR peaks at different subgap energy \cite{Ruby2016,Choi2017}. 

\begin{figure*}[b]
  \includegraphics[width=0.6\textwidth]{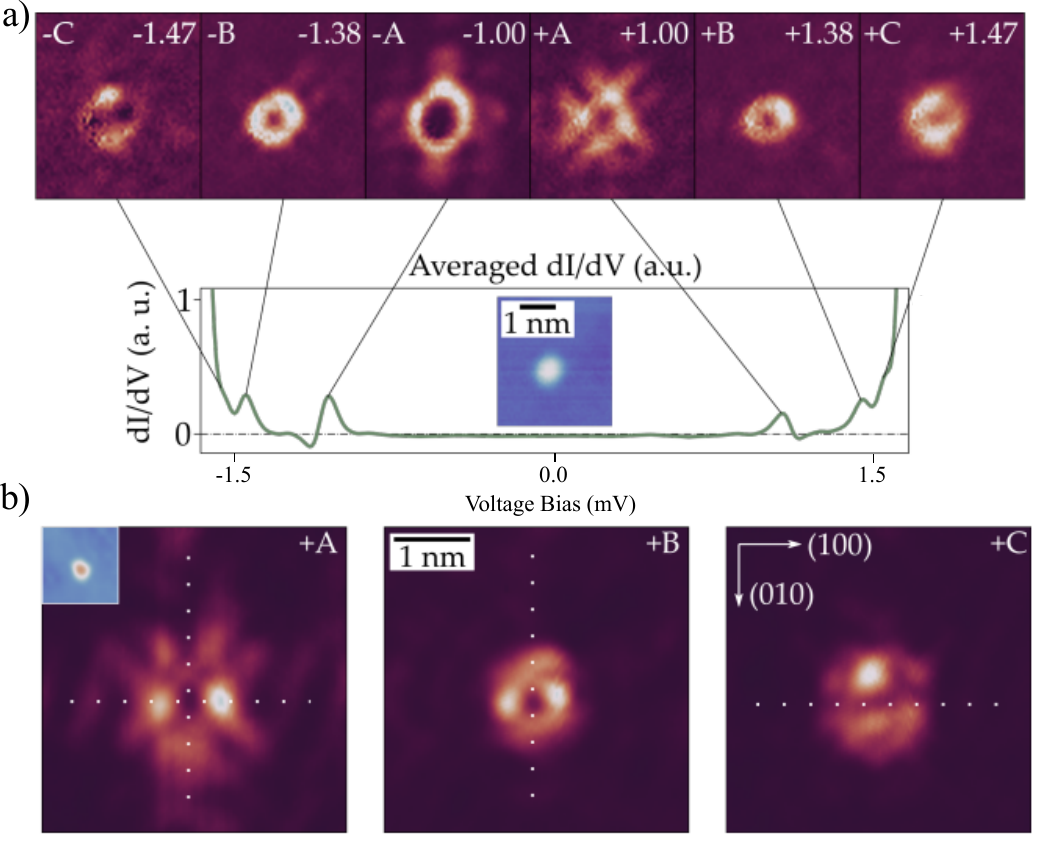}
  \caption{Spatial distribution of YSR states over a V Adatom: 
  a) Conductance maps at the energies of each YSR resonance shown in the spectra. The maps were constructed from constant bias cuts extracted from a 56x56 grid of dI/dV spectra measured over the 3x3-nm$^2$ area in the blue image at the inset. The different shapes of particle-hole symmetry YSR distributions were explained in \cite{Choi2017}.
  b) Dual-pass dI/dV maps measured over a different V adatom than in panel a), using a different microscopic tip. The three dI/dV maps were done by mapping dI/dV signal (V$_{rms}$=50 $\mu$V) at the (subgap) positive bias values of the three YSR states, following a tip trajectory measured in a previous line pass with V=-3 mV, I=300 pA. 
  Pointed lines indicate the symmetry axis in each map. These maps allow us to interpret the three peaks as three independent  YSR channels caused by the interaction of the three spin-polarized d orbitals of V with the \bipd\  substrate, hinting that V adatoms maintain their S=3/2 state when adsorbed on the surface.
  } \label{SF-Orbital}
\end{figure*}

To probe the orbital origin of the three YSR peaks observed in STS spectra, we mapped the spatial distribution of the YSR amplitude around an V adatom. As shown in Figure~\ref{SF-Orbital},  the YSR signal over the V adatoms follows a pattern of lobes and nodal planes that resembles the amplitude of three different $3d$-orbitals, demonstrating their orbital-specific origin \cite{Ruby2016,Choi2017}. This proves that the $3d_3$ valence state of V survives on the surface. Therefore, the three YSR states correspond to three orbital channels, each accounting for the hybridization of a V $d$ orbital with a substrate band. The same extension is observed for the three sub-gap excitations (Fig. \ref{SF:QPI1}c), so we assume that the three orbitals of Vanadium are coupled to the same surface band. Other surface and bulk-projected bands with little or no hybridization with the impurity remain unperturbed.

 \newpage

\section{Bogoliuvov quasiparticle interference of YSR states} 
 \label{SM:QPI-allYSR}

\begin{figure*}[hb]
  \includegraphics[width=0.8\textwidth]{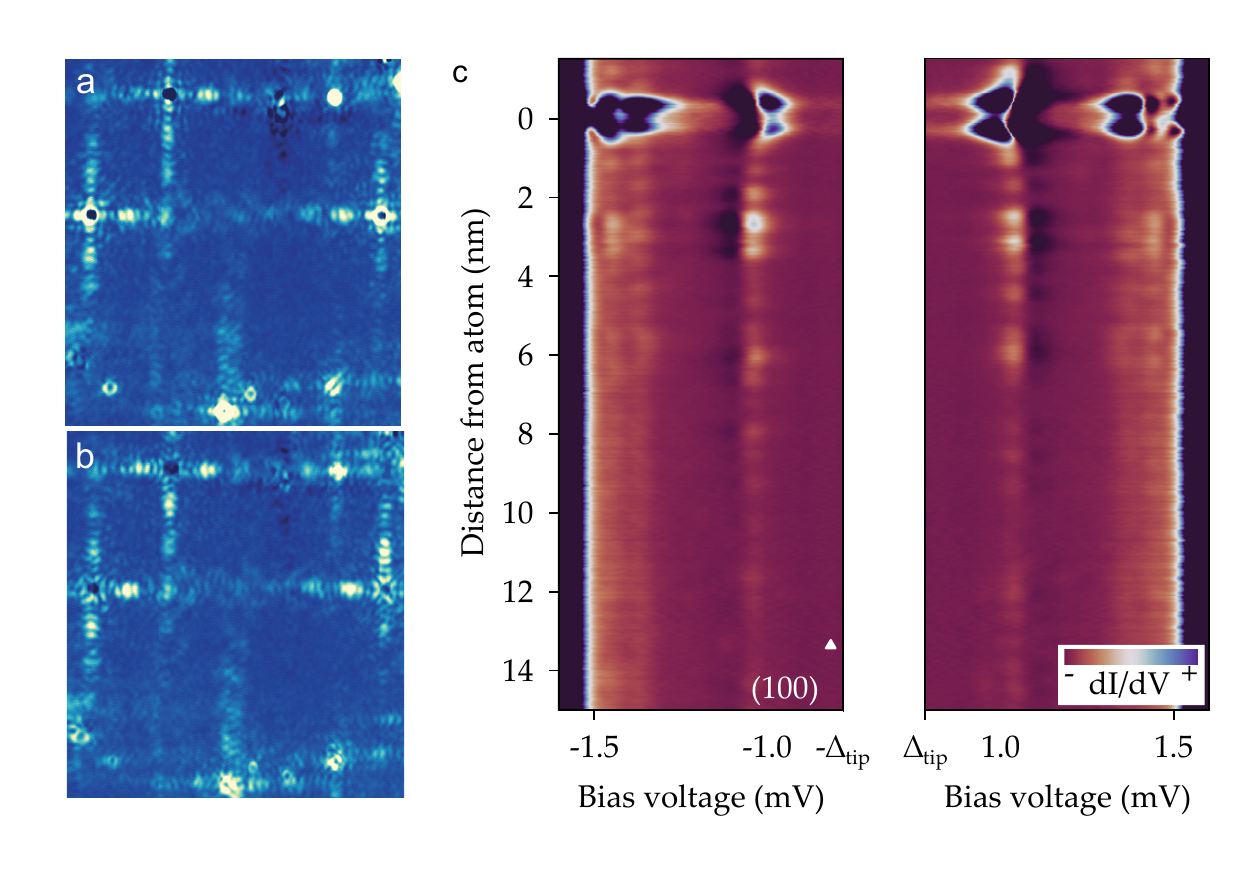}
  \caption{a, b) 30~nmx30~nm dI/dV maps measuring at the bias value of the particle and hole components of the $\alpha$ YSR resonance, respectively. The maps are obtained using a dual pass technique, described above. c) Spectral line maps of subgap dI/dV plots measured at increasing distances from a V adatom along a high-symmetry direction of the substrate, i.e., along a YSR beam as shown in panels a) and b). } \label{SF:QPI1}
\end{figure*}

Figures \ref{SF:QPI1}a,b) compares the dI/dV map of the BQPI pattern in Fig. 2 of the manuscript at both polarities. Similar oscillations are observed for all YSR states (see \ref{SF:QPI1}c)), however, the intensity  of the YSR oscillations reduces as the YSR states are closer to the superconducting gap of the bulk substrate.

 The wavefunction of Shiba states contains two sinusoidal terms which may produce an oscillation:

\begin{equation}
\psi_\pm (r) = \frac{1}{\sqrt{N}}\frac{\sin(k_Fr+\delta^\pm)}{k_Fr}e^{-\Delta sin(\delta^+-\delta^-)r/\hbar v_F}
\label{eq:shiba}
\end{equation}

where $\pm$ refers to the positive and negative component of the Shiba peak, $N$ is the density of states in the metallic state, $k_F$ is the Fermi wavector, $\delta$ the phase difference, and $v_F$ is the Fermi velocity. We can rule out the second sinusoidal term because the phase difference is constant for a given energy~\cite{Heinrich2018} and thus all modulating behavior can be traced back to the Fermi vectors in $\beta$-Bi$_2$Pd. Under this condition, the modulations will resemble the behavior of Friedel oscillations in conventional metals, what allows for the interpretation of the modulations in the superconducting state in terms of those in the normal metal.

 
\newpage



 

\newpage

\section{Helical multi-band superconductor}\label{SN:theo2}

In this section, we present a theoretical model that explains the main features of the Bogoliubov QPI (BQPI) pattern observed in the experiment. These features are:
\begin{enumerate}
    \item A single impurity interacting with a helical band does not generate any BQPI pattern.\label{point1}

    \item A single impurity coupled to a helical band does not generate any BQPI pattern, even if an interband scattering term is present.\label{point2}

    \item Interband hybridization is required to account for the complex BQPI patterns observed in our system, not a scattering term.\label{point3}
    
\end{enumerate}

We describe the system superconductor plus impurity by the  following effective Hamiltonian:
\begin{equation}
\hat H=\hat H_{0}+\hat H_{imp}\label{eq:Hamiltonian}\;
\end{equation}
where, $\hat H_{0}$ describes the superconducting host consisting of two helical bands. 
Written in the
spin-Nambu-band basis ($\sigma-\tau-\eta)$ it reads: 
\begin{equation}
\check H_{0p}=\left(\begin{array}{cccc}
 pv_{1}\check\sigma_{z}-\mu & \Delta & t & 0\\
\Delta & - pv_{1}\check\sigma_{z}+\mu & 0 & -t\\
t & 0 &  pv_{2}\check\sigma_{z}-\mu & \Delta\\
0 & -t & \Delta & - pv_{2}\check\sigma_{z}+\mu
\end{array}\right)\label{eq:H0}
\end{equation}

Or equivalently: 
\begin{equation}
\check H_{0p}=( pv_{1}\check\sigma_{z}-\mu)\check\tau_{3}\frac{\check\eta_{3}+\check\eta_{0}}{2}+(\hat p v_{2}\check\sigma_{z}-\mu)\check\tau_{3}\frac{\check\eta_{0}-\check\eta_{3}}{2}+\Delta\check\tau_{1}+t\check\tau_{3}\check\eta_{1}\label{eq:H0_2}
\end{equation}
where, $p$ is the momentum,  $v_i$ ($i=1,2$) are the Fermi velocities of each band, $\Delta$ is the order parameter (equal for both bands, as we measure a single gap), $\mu$ is the chemical potential and $t$ is an interband hopping term, responsible for band hybridization. The $\check{ }$ symbol refers to matrices. For simplicity, we use a 1D Hamiltonian to describe the system. This approximation is justified by the fact that the focusing effect arising from the square-shaped Fermi contour leads to a spatial dependence of the YSR state reminiscent of a 1D situation Ref. \cite{ortuzar2022}.  Our 1D Hamiltonian, Eq. (\ref{eq:H0_2}), will describe the spatial dependence of the subgap states along the crystallographic directions (100) and (010). 

Within our model, the impurity is localized at $x=0$. The Hamiltonian describing the exchange coupling between the impurity and the free electrons reads
\begin{equation}\label{eq:Himp}
\check H_{imp}=\vec{h}\cdot\check{\vec{\sigma}}\delta(x)\check\tau_{0}\check\eta_{0}\;.
\end{equation}
\\

\textbf{1 - A single impurity interacting with a helical band does not generate any BQPI pattern:}
We start by considering two decoupled bands, i.e. $t=0$ in Eq. \eqref{eq:H0}. We  solve for one of the spin directions ($v_i\rightarrow -v_i$ flips the spin):
\begin{equation}
    \check H_{0i}=(v_i\check \sigma_z p-\mu)\check \tau_3+\Delta \check \tau_1
\end{equation}
the Green Function (GF) in momentum space reads
\begin{equation}\label{GFhel}
    \check G_{0}(\omega,p)=[\omega\mathcal{I}-\check H_{0p}]^{-1}=\dfrac{1}{(\Delta^2-\omega^2)+(pv_i\check\sigma_z-\mu)^2}(\omega+(p v_i \check\sigma_z -\mu)\check\tau_3+\Delta\check\tau_1)
\end{equation}
we get the real space GF from the FT of this equation. In a 1D system, one can always solve the integral by finding the poles in the positive or negative imaginary plane. The poles read $p v_i=\sigma(\pm i\Omega+\mu)$, with $\Omega=\sqrt{\Delta^2-\omega^2}$. For $x>0$ we look for poles in the upper half-plane and for $x<0$ in the lower one. 
The real space GF reads:
\begin{equation}
    \check G_{0}(x)=\dfrac{-1}{v_i}(\dfrac{\omega}{\Omega}+i\text{sign}(x)\check\tau_3+\dfrac{\Delta}{\Omega}\check\tau_1)e^{-i x\check\sigma_z\mu/v_i-\Omega|x|/v_i}\;. 
\end{equation}
 The GF describing the YSR state can be found from the close expression
\begin{equation}\label{eq:full_sol}
\check G(x,x)=\check G_{0}(0)+\check G_{0}(x)\vec{h}\vec{\sigma}\left[\check{1}-\check G_{0}(0)\vec{h}\check{\vec{\sigma}}\right]^{-1}\check G_{0}(-x)\; .
\end{equation}
The YSR states arise from the poles of the T-matrix ($\check T=\vec{h}\cdot\check{\vec{\sigma}}\left[\check{1}-\check G_{0}(0)\vec{h}\cdot\check{\vec{\sigma}}\right]^{-1}$)\cite{wang2003,bena2005,balatsky2006}, and their spatial extension is defined by the second term in  Eq. \eqref{eq:full_sol}. The T-matrix is multiplied by two GFs with opposite sign position inputs ($\check G_0(x)$ and $\check G_0(-x)$). These two present reversed oscillatory behavior ($e^{-i x\check\sigma_z\mu/v_i}$ and $e^{i x\check\sigma_z\mu/v_i}$) so that they cancel each other upon multiplication \footnote{We are only considering the trace of Eq. \eqref{eq:full_sol}, proportional to the DoS, the magnitude that we measure. Non-diagonal terms can oscillate.}. This leaves no oscillations in the second term of Eq. \eqref{eq:full_sol}, i.e., no oscillations of the extension of the DoS of a YSR state in a helical system. This can be easily understood by noting that each spin band only cuts the Fermi level once in the helical limit, so an impurity connected to this band can not create any BQPI, as it can not be scattered to any other point in the Fermi contour. With this, we prove our first statement.

\textbf{2 - A single impurity interacting with a helical band with an interband scattering term  to other bands does not generate any BQPI pattern.}
We again assume two independent bands,  but assume that the impurity has an interband scattering term:
\begin{align}
    &\check H_0=\begin{pmatrix}
        \check H_1 & 0\\
        0 &\check H_2
    \end{pmatrix} & &\check V=\check H_{imp}=\vec{h}\cdot\check{\vec{\sigma}}\delta(x)\check\tau_{0}\check\eta_{1}=\begin{pmatrix}
        0 & \vec{h}\cdot\check{\vec{\sigma}}\\
        \vec{h}\cdot\check{\vec{\sigma}} & 0
    \end{pmatrix}\delta(x)\; .
\end{align}
The GF of the bare superconductor will be block diagonal and each block will have the same form as in eq. \eqref{GFhel}, where $v_F$ is exchanged by $v_i$. Taking the Dyson equation for one of the bands we get
\begin{equation}
     G_{ii}(x,x)=G_{0ii}(0)+G_{0ij}(x)[\mathcal{I}-G(0)V]_{jk}G_{0ki}(-x)=G_{0ii}(0)+G_{0ii}(x)[\mathcal{I}-G(0)V]_{0ii}G_{0ii}(-x)\; ,
\end{equation}
where we used $G_{0ij}=G_{0}\delta_{ij}$. We again note that the only terms that will oscillate are the nondiagonal terms of $G_{ij}$, but these terms do not contribute to the LDoS, defined as $\rho\propto \Tr\Im(G(x))$. Then, a single impurity - even with an interband scattering term - will not lead to any spatial oscillatory behavior in a multiband helical system, proving our second statement. 
\\

\textbf{3 - Interband hybridization is required to account for the complex BQPI patterns observed in our system:} 
Finally, we solve the full Hamiltonian [Eq. \eqref{eq:H0}] with a finite hopping term $t$. The poles of the GF now read,
\begin{equation}
    p=\dfrac{-V(\mu\pm i \Omega\pm) \sqrt{t^2(V^2-v^2)+(\mu \pm i\Omega)^2}}{V^2-v^2}
\end{equation}
with $V=v_1+v_2$ and $v=v_1-v_2$. Depending on the relative spin orientation of the bands, $V$ and $v$ can be either positive or negative. The GF in real space then reads:
\begin{equation}
    G(x)= e^{(i\mu-\Omega|x|)/v_1}\dfrac{\omega}{v_1\Omega}+e^{(i\mu-\Omega|x|)/v_2}\dfrac{\omega}{\Omega}\dfrac{t^2 v_2}{(v_1^2-v_2^2)\mu^2}+\mathcal{O}\left(\dfrac{t}{\mu}\right)^3
\end{equation}
One can check from the  Dyson equation that the diagonal terms have an oscillatory behavior arising from the scattering between the two original bands: $\exp{ix\mu(1/v_1+1/v_2)}=\exp{ix(p_{F1}+p_{F2})}$. The intensity of these oscillations is $\mathcal{O}(t/\mu)^2$.

\begin{figure*}
    \includegraphics[width=0.99\textwidth]{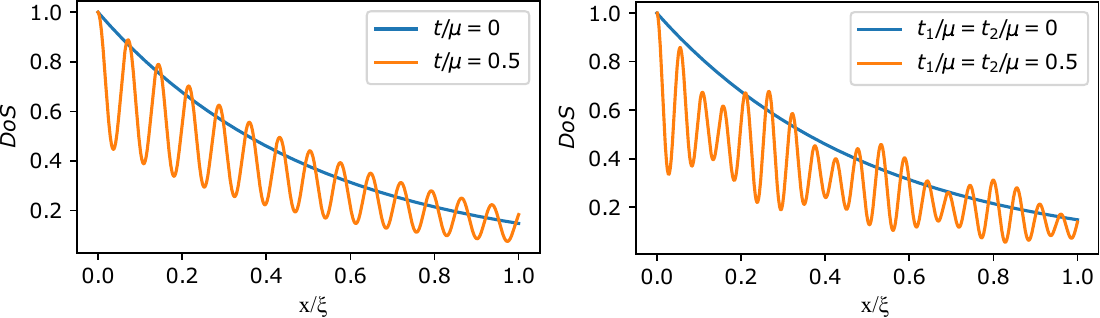}
    \caption{Calculated extension of the YSR state for  (a) a two-band model ($v_{F1}=0.1$, $v_{F1}=0.25$ and $\Delta/\mu=0.01$) and (b) a three-band model ($v_{F1}=0.1$, $v_{F1}=0.25$, $v_{F3}=0.6$ and $\Delta/\mu=0.01$). The horizontal axis is normalized to the coherence length of the superconductor.}
    \label{fig:2osc}
\end{figure*}

Fig. \ref{fig:2osc} shows the results obtained from the effective helical multiband Hamiltonian. Fig. \ref{fig:2osc} (a) compares the evolution of a YSR state for a Hamiltonian without hopping (blue) and one with finite hopping between the bands (orange). In the latter case, an oscillatory behavior can be seen, arising from the combination of the two Fermi momenta of the bands.  The oscillation period is not given by $k_{F1}+k_{F2}$, but a more complex function of $t$: $k_{F1}+k_{F2}+\mathcal{O}(t/\mu)^2$. To get a double period oscillation, we need to add a third band. This scenario is depicted in Fig. \ref{fig:2osc} (b). The oscillatory behavior is again compared with the no-hopping limit. 
Thus, according to our model, the multiple-period oscillations can only be explained by assuming a hybridization between the bands.

\section{The origin of the hybridization} \label{SN:theo3}

As demonstrated in the previous analysis, single adatoms, due to their localized nature, do not produce oscillations from interband scattering. This concerns both BQPI and QPI. However, quasiparticle interference (QPI) patterns have been extensively used to probe the Fermi surface in systems with spin-orbit coupling.  In most of these cases, the studied bands are Rashba-split bands, and hence, are not orthogonal \cite{Kohsaka2017}. 

The observation of all possible scattering vectors in the normal state QPI (e.g. as in Fig. S2) suggests that all bands at the \bipd\ surface are coupled.  DFT simulations \cite{Sakano2015} find that the bands overlap in momentum space far from \ef, and, thus, it is expected that they easily hybridize around \ef. This points to an intrinsic hybridization mechanism, of the bands. 
 
Here, we propose an alternative possibility: a finite distribution of scatterers on a 2D surface, such as an ensemble of adatoms, vacancies, or subsurface impurities, as it is the case in the as-cleaved \bipd\ surface, can behave as an effective inter-band coupling term and allow for inter-band QPI \cite{Sharma2020,dutt2017,korshunov2016}. With this description, the inter-band scattering can be understood as a multi-impurity scattering path. 


We calculate the disorder-averaged GF of a multiband metal with impurities that can connect different bands. The scattering potential reads,
\begin{equation}
    \hat U=\sum_{\sigma\alpha\beta i}u^{\alpha\beta}\hat\psi^{\dagger}_{\alpha}(r_i)\hat\psi_{\beta}(r_i)
\end{equation}
$\alpha=\beta$ is the intraband scattering while $\alpha\neq \beta$ is the interband scattering. Assuming a Gaussian distribution of the impurity positions, only two point impurity vertices will contribute to the total self energy of the disordered GF:
\begin{equation}
    \langle u(r)u(r')\rangle=\dfrac{1}{2\pi\nu\tau}\delta(r-r')
\end{equation}
so that $ \langle u_qu_{q'}\rangle=\dfrac{1}{2\pi\nu\tau L^d}\delta_{q+q',0}$. The self energy reads,
\begin{equation}
    \Sigma=\dfrac{1}{2\pi\nu\tau L^d}\sum_q \hat{U}G_{k+q}\hat{U} \text{, with } \hat{U}=u\check\eta_0+v\check\eta_1
\end{equation}
The $\check\eta_i$ are the band space Pauli matrices. The disorder averaged GF reads
\begin{equation}\label{avGF}
\begin{split}
    \langle G_k \rangle&=\left(1- G_k\Sigma\right)^{-1}G_k\\
    &=\begin{pmatrix}
        \dfrac{\xi_b-\omega+\Sigma_{bb}}{|\Sigma_{ab}|^2-(\xi_b-\omega+\Sigma_{bb})(\xi_a-\omega+\Sigma_{aa})} & \dfrac{\Sigma_{ab}}{|\Sigma_{ab}|^2-(\xi_b-\omega+\Sigma_{bb})(\xi_a-\omega+\Sigma_{aa})}\\
        \dfrac{\Sigma_{ba}}{|\Sigma_{ab}|^2-(\xi_b-\omega+\Sigma_{bb})(\xi_a-\omega+\Sigma_{aa})}& \dfrac{\xi_a-\omega+\Sigma_{aa}}{|\Sigma_{ab}|^2-(\xi_b-\omega+\Sigma_{bb})(\xi_a-\omega+\Sigma_{aa})}
    \end{pmatrix}
\end{split}
\end{equation}
The impurity averaged GF has the same form that the GF one would get from a two-band Hamiltonian with an interband hopping term.  $t=\Sigma_{ab}\sim 1/\tau$, where $l=v_F\tau$ is the mean free path of the electrons. For $t$ to be of order $\mathcal{O}(E_F/10)$, $l\sim$10~nm

%